\input harvmac.tex

\input epsf.tex

\def\figin{\epsfcheck\figin}\def\figins{\epsfcheck\figins}
\def\epsfcheck{\ifx\epsfbox\UnDeFiNeD
\message{(NO epsf.tex, FIGURES WILL BE IGNORED)}
\gdef\figin##1{\vskip2in}\gdef\figins##1{\hskip.5in}
\else\message{(FIGURES WILL BE INCLUDED)}%
\gdef\figin##1{##1}\gdef\figins##1{##1}\fi}
\def\DefWarn#1{}
\def\figinsert{\goodbreak\midinsert}
\def\ifig#1#2#3{\DefWarn#1\xdef#1{fig.~\the\figno}
\writedef{#1\leftbracket fig.\noexpand~\the\figno}%
\figinsert\figin{\centerline{#3}}\medskip\centerline{\vbox{\baselineskip12pt
\advance\hsize by -1truein\noindent\footnotefont{\bf Fig.~\the\figno:} #2}}
\bigskip\endinsert\global\advance\figno by1}

\font\sf = cmss10


\lref\MaldacenaIM{ J.~M.~Maldacena, ``Wilson loops in large N
field theories,'' Phys.\ Rev.\ Lett.\  {\bf 80}, 4859 (1998)
[arXiv:hep-th/9803002].
}

\lref\ReyIK{ S.~J.~Rey and J.~T.~Yee, ``Macroscopic strings as
heavy quarks in large N gauge theory and  anti-de Sitter
supergravity,'' Eur.\ Phys.\ J.\ C {\bf 22}, 379 (2001)
[arXiv:hep-th/9803001].
}

\lref\BrandhuberER{ A.~Brandhuber, N.~Itzhaki, J.~Sonnenschein and
S.~Yankielowicz, ``Wilson loops, confinement, and phase
transitions in large N gauge  theories from supergravity,'' JHEP
{\bf 9806}, 001 (1998) [arXiv:hep-th/9803263].
}

\lref\KruczenskiUQ{ M.~Kruczenski, D.~Mateos, R.~C.~Myers and
D.~J.~Winters, ``Towards a holographic dual of large-N(c) QCD,''
arXiv:hep-th/0311270.
}

\lref\KarchSH{ A.~Karch and E.~Katz, ``Adding flavor to AdS/CFT,''
JHEP {\bf 0206}, 043 (2002) [arXiv:hep-th/0205236].
}

\lref\KruczenskiBE{ M.~Kruczenski, D.~Mateos, R.~C.~Myers and
D.~J.~Winters, ``Meson spectroscopy in AdS/CFT with flavour,''
JHEP {\bf 0307}, 049 (2003) [arXiv:hep-th/0304032].
}

\lref\BachasXS{  C.~Bachas, ``Convexity Of The Quarkonium
Potential,'' Phys.\ Rev.\ D {\bf 33}, 2723 (1986).
}

\lref\ChernodubBK{ M.~N.~Chernodub, F.~V.~Gubarev,
M.~I.~Polikarpov and V.~I.~Zakharov, ``Confinement and short
distance physics,'' Phys.\ Lett.\ B {\bf 475}, 303 (2000)
[arXiv:hep-ph/0003006].
}

\lref\PinedaSE{ A.~Pineda, ``The static potential: Lattice versus
perturbation theory in a renormalon-based approach,'' J.\ Phys.\ G
{\bf 29}, 371 (2003) [arXiv:hep-ph/0208031].
}

\lref\MaldacenaRE{
J.~M.~Maldacena,
``The large N limit of superconformal field theories and supergravity,''
Adv.\ Theor.\ Math.\ Phys.\  {\bf 2}, 231 (1998)
[Int.\ J.\ Theor.\ Phys.\  {\bf 38}, 1113 (1999)]
[arXiv:hep-th/9711200].
}

\lref\witten{E.~Witten,
``Anti-de Sitter space, thermal phase transition, and confinement in  gauge
theories,''
Adv.\ Theor.\ Math.\ Phys.\  {\bf 2}, 505 (1998)
[arXiv:hep-th/9803131].}

\lref\gross{D.~J.~Gross and H.~Ooguri,
``Aspects of large N gauge theory dynamics as seen by string theory,''
Phys.\ Rev.\ D {\bf 58}, 106002 (1998)
[arXiv:hep-th/9805129].}

\lref\hashi{A.~Hashimoto and Y.~Oz,
``Aspects of {QCD} dynamics from string theory,''
Nucl.\ Phys.\ B {\bf 548}, 167 (1999)
[arXiv:hep-th/9809106].
}

\lref\barbon{J.~L.~F.~Barbon, C.~Hoyos, D.~Mateos and R.~C.~Myers,
``The holographic life of the eta',''
arXiv:hep-th/0404260.
}

\lref\Kruczenskibe{ M.~Kruczenski, D.~Mateos, R.~C.~Myers and
D.~J.~Winters, ``Meson spectroscopy in AdS/CFT with flavour,''
JHEP {\bf 0307}, 049 (2003) [arXiv:hep-th/0304032].}

\lref\Myersps{ R.~C.~Myers, ``Dielectric-branes,'' JHEP {\bf
9912}, 022 (1999) [arXiv:hep-th/9910053].}

\lref\Myersbw{ R.~C.~Myers, ``Nonabelian phenomena on D-branes,''
Class.\ Quant.\ Grav.\  {\bf 20}, S347 (2003)
[arXiv:hep-th/0303072].}

\lref\GubserTV{ S.~S.~Gubser, I.~R.~Klebanov and A.~M.~Polyakov,
``A semi-classical limit of the gauge/string correspondence,''
Nucl.\ Phys.\ B {\bf 636}, 99 (2002) [arXiv:hep-th/0204051].
}

\lref\KarchXE{ A.~Karch, E.~Katz and N.~Weiner, ``Hadron masses
and screening from AdS Wilson loops,'' Phys.\ Rev.\ Lett.\  {\bf
90}, 091601 (2003) [arXiv:hep-th/0211107].
}

\lref\MaldacenaRE{ J.~M.~Maldacena, ``The large N limit of
superconformal field theories and supergravity,'' Adv.\ Theor.\
Math.\ Phys.\  {\bf 2}, 231 (1998) [Int.\ J.\ Theor.\ Phys.\  {\bf
38}, 1113 (1999)] [arXiv:hep-th/9711200].
}

\lref\BaliVR{ G.~S.~Bali {\it et al.}  [TXL Collaboration],
``Static potentials and glueball masses from QCD simulations with
Wilson  sea quarks,'' Phys.\ Rev.\ D {\bf 62}, 054503 (2000)
[arXiv:hep-lat/0003012].
}

\lref\PeetWN{ A.~W.~Peet and J.~Polchinski, ``UV/IR relations in
AdS dynamics,'' Phys.\ Rev.\ D {\bf 59}, 065011 (1999)
[arXiv:hep-th/9809022].
}

\lref\KlebanovHB{ I.~R.~Klebanov and M.~J.~Strassler,
``Supergravity and a confining gauge theory: Duality cascades and
chiSB-resolution of naked singularities,'' JHEP {\bf 0008}, 052
(2000) [arXiv:hep-th/0007191].
}

\lref\ItzhakiDD{ N.~Itzhaki, J.~M.~Maldacena, J.~Sonnenschein and
S.~Yankielowicz, ``Supergravity and the large N limit of theories
with sixteen  supercharges,'' Phys.\ Rev.\ D {\bf 58}, 046004
(1998) [arXiv:hep-th/9802042].
}

\lref\SusskindDQ{ L.~Susskind and E.~Witten, ``The holographic
bound in anti-de Sitter space,'' arXiv:hep-th/9805114.
}

\lref\ChodosGT{ A.~Chodos and C.~B.~Thorn, ``Making The Massless
String Massive,'' Nucl.\ Phys.\ B {\bf 72}, 509 (1974).
}

\lref\FramptonWV{ {\sl see for instance} P.~H.~Frampton, ``Dual
Resonance Models And Superstrings, W. A. Benjamin (1974)''
}

\lref\BigazziZE{ F.~Bigazzi, A.~L.~Cotrone, L.~Martucci and
L.~A.~Pando Zayas, ``Wilson loop, Regge trajectory and hadron
masses in a Yang-Mills theory from semiclassical strings,''
arXiv:hep-th/0409205.
}

\lref\AbrikosovSX{ A.~A.~Abrikosov, ``On The Magnetic Properties
Of Superconductors Of The Second Group,'' Sov.\ Phys.\ JETP {\bf
5}, 1174 (1957) [Zh.\ Eksp.\ Teor.\ Fiz.\ {\bf 32}, 1442 (1957)].
}

\lref\BabingtonVM{ J.~Babington, J.~Erdmenger, N.~J.~Evans,
Z.~Guralnik and I.~Kirsch, ``Chiral symmetry breaking and pions in
non-supersymmetric gauge /  gravity duals,'' Phys.\ Rev.\ D {\bf
69}, 066007 (2004) [arXiv:hep-th/0306018].
}

\lref\NielsenCS{ H.~B.~Nielsen and P.~Olesen, ``Vortex-Line Models
For Dual Strings,'' Nucl.\ Phys.\ B {\bf 61}, 45 (1973).
}

\lref\BachasFR{ C.~Bachas and M.~Petropoulos, ``Anti-de-Sitter
D-branes,'' JHEP {\bf 0102}, 025 (2001) [arXiv:hep-th/0012234].
}

\lref\KarchGX{ A.~Karch and L.~Randall, ``Open and closed string
interpretation of SUSY CFT's on branes with boundaries,'' JHEP
{\bf 0106}, 063 (2001) [arXiv:hep-th/0105132].
}

\lref\ZhangCD{ A.~Zhang, ``Regge trajectories analysis to
D*/sJ(2317)+-, D/sJ(2460)+- \ \ and D/sj(2632)+ mesons,''
arXiv:hep-ph/0408124.
}

\lref\Glozman{ L. Ya. Glozman, ``Chiral and U(1)$_A$ restorations
high in the hadron spectrum, semiclassical approximation and large
Nc,'' arXiv:hep-ph/0411281.
}


{ \Title{\vbox{\baselineskip12pt \hbox{hep-th/0412260 }
{\vbox{\baselineskip12pt \hbox{CPHT-RR 069.1204 } }} }}
 {\vbox{
{\centerline { Multiflavour excited mesons  from the fifth
dimension}
}}}}

\bigskip
\centerline{ Angel Paredes$^a$ and Pere Talavera$^b$ }
\bigskip~

\centerline{$^a$Centre de Physique Th{\'e}orique,  Ecole
Polytechnique,} \centerline{ 91128 Palaiseau, France}

\medskip

\centerline{$^b$ Departament de F{\'\i}sica i Enginyeria Nuclear,}
\centerline{ Universitat Polit\`ecnica de Catalunya, Jordi Girona
1--3, E-08034 Barcelona, Spain }

\vskip .3in

\baselineskip12pt

\vfill

We study the Regge trajectories and the quark-antiquark energy in
excited  hadrons composed by different dynamical mass constituents
via the gauge/string correspondence. First we exemplify the
procedure in a supersymmetric system, D3-D7, in the extremal case.
Afterwards we discuss the model dual to large-$N_c$ QCD, D4-D6
system. In the latter case we find the field theory expected gross
features of vector like theories: the spectrum resembles that of
heavy quarkonia and the Chew-Frautschi plot of the singlet and
first excited states is in qualitative agreement with those of
lattice QCD. We stress the salient points of including different
constituents masses.

\Date{December  2004} \eject \baselineskip14pt

\newsec{Motivation}

One of the main features of vector-like theories is the
self-interaction of the spin one gauge fields. With the inclusion
of fundamental matter, colour charged,  two competing
contributions appears to the quark antiquark potential: one,
attractive, is due to the interchange of the gauge bosons while
the second, repulsive, comes from the self-interaction of these
gauge bosons. The net effect is the formation of an electric flux
between the two colour sources. The confined, electric flux tube
connecting massless quarks can be modelled by a string, which
represents one of the  simplest hadron models
\refs{\AbrikosovSX,\NielsenCS}. When the string tension is
constant it leads to a simple relation between the energy and the
spin of the string, $ E \sim \sqrt{J}$, {\sl i.e.} Regge. {F}rom
the quantum states assignment, Regge trajectories are based on
flavour independence arguments and exchange degeneracy, {\sl i.e.}
$J^{PC}$ and $(J+1)^{-P,-C}$ states belong to a common Regge
trajectory. Even though it is well known that more realistic Regge
behaviours include a positive curvature and a nonzero intercept
\ZhangCD.

\medskip

It is our aim to recast the features of these models in the
gravity/field theory duality framework. {F}or that purpose we
shall deal with the meson Regge trajectory in the light of
holographic duals of confining supergravity backgrounds. In the
first holographic setup \refs{\ReyIK,\MaldacenaIM} the addition of
matter was achieved by higgsing the initial gauge group: A stack
of $N$ D3-branes was considered at the origin giving rise to a
gauge group $SU(N)$. Then one of the branes was pulled out up to
the boundary breaking the initial gauge group to $SU(N-1)\times
U(1)$. The string length stretching from the origin up to the
boundary is proportional to the mass of the quark attached at one
of the string ends. Then the final picture was an infinite heavy
quark at the boundary. Considering two of such strings a more
stable configuration, merging the lower end points, is possible
giving a bound state of two infinite masses.

\medskip

The addition of finite flavoured matter is achieved by means of an
open string sector \KarchSH. The end points of these are supported
by $N_f$ nonbackreacting filling space-time D-branes, $N_f\ll
N_c\rightarrow \infty\,.$ In most of the cases, D-branes carry
RR-gauge fields and hence a nonzero charge. A possible mechanism
to stabilise them at finite distance is by wrapping trivial cycles
\refs{\BachasFR,\KarchGX}. Then, without creating instabilities,
negative massive modes prevent the slipping of the D-brane out of
the wrapped trivial cycle. The final result are a set of mesons
(in the large-$N_c$ limit) consisting of bound state poles.
Neglecting meson widths correspond to disregard mixing between
$q\bar{q}$ and other sectors of the theory as glueballs or
$qq\bar{q}\bar{q}\,.$

The meson spectrum can be divided in two different sectors: {\sl
i)} A sector with small ($J=0,1,2$) angular momentum, that can be
tackled considering small fluctuations of the probe-brane. {\sl
ii)} Stringy states at large-$J$ corresponding to excited meson
states. In the latter case they are represented by rotating
Wilson-loops. Notice that this inherently implies a semi-classical
limit. It is then natural to assume that isolated hadrons exist at
sufficient energy to be treated semi-classically \Glozman .

\medskip

We shall consider the spectrum of the stringy states in a generic
case: an hadron formed by two different constituents masses. This
will allow us to study some features of the hadron excited spectra
as a function of the masses in a more realistic set up. There is,
for instance, little knowledge of heavy-light charmed mesons.
While at low-energy there is no significant change with respect to
the degenerate case, at high-energy the entire profile changes. In
the cases where the metric has been obtained via compactification
we argue that in the vicinity of the high-energy region there is
the opening of the compactified dimensions, and hence the dual
field theory is changed.

In order to introduce notation and the main work line we deal with
an extremal supersymmetric model, D3-D7. We discuss the spectrum,
the energy-momenta relation and the quark-antiquark energy. We pay
especial attention to the role of mass subtraction. We turn then
to evaluate the above features in a nonsupersymmetric model,
D4-D6, comparing their qualitative behaviour with the experimental
ones. In addition we comment on the quark-antiquark potential
screening and on the 't Hooft line.

\lref\NunezCF{ C.~Nunez, A.~Paredes and A.~V.~Ramallo, ``Flavoring
the gravity dual of N = 1 Yang-Mills with probes,'' JHEP {\bf
0312}, 024 (2003) [arXiv:hep-th/0311201].
}

\lref\OuyangDF{ P.~Ouyang, ``Holomorphic D7-branes and flavored N
= 1 gauge theories,'' Nucl.\ Phys.\ B {\bf 699}, 207 (2004)
[arXiv:hep-th/0311084].
}

\lref\AreanMM{ D.~Arean, D.~Crooks and A.~V.~Ramallo,
``Supersymmetric probes on the conifold,'' JHEP {\bf 0411}, 035
(2004) [arXiv:hep-th/0408210].
}

\lref\KupersteinHY{ S.~Kuperstein, ``Meson spectroscopy from
holomorphic probes on the warped deformed conifold,''
arXiv:hep-th/0411097.
}

Even if we treated ${\cal N}=2$~ and ${\cal N}=0$~ theories it
would also be interesting to perform a similar analysis for ${\cal
N}=1$ theories although, being confining, it is natural to expect
a similar behaviour to that of the non-supersymmetric case. The
embedding of the corresponding flavour branes was studied \NunezCF
in the case of a non-conformally flat setup and in
\refs{\OuyangDF,\AreanMM,\KupersteinHY} for the theories on the
conifold.

\lref\ChodosGT{ A.~Chodos and C.~B.~Thorn, ``Making The Massless
String Massive,'' Nucl.\ Phys.\ B {\bf 72}, 509 (1974).
}

\newsec{The $D3-D7$ brane model}

We start with a supersymmetric system that allows a fairly good
analytic control over the full calculation.  To obtain a system
with matter in the bi-fundamental representation we make use of
the observation that a holographic dual to $4$-dimensional
super-Yang-Mills (SYM) theory can be obtained in the near-horizon
limit of the D3-D7 system \KarchSH.

The set up consists in a stack of $N_c$ D3-branes at the origin
expanding in four-directions ($0,1,2,3$) and warping along the
transverse ones. This give rise to an ${\cal N}=4$ SYM field
theory. Another set of $N_f$ D7-branes are located along
($0,1,2,3,4,5,6,7$) and warping in the rest of dimensions. The
insertion of this last stack of branes breaks the initial ${\cal
N}=4$ symmetry to ${\cal N}=2$ SYM at low-energy. We shall
consider the case $N_f \ll N_c$, then the D7-brane backreaction
can be ignored.

\medskip

The D3-brane system can be obtained from type-IIB string theory on
the $AdS_5 \times S^5$ background
\eqn\ads{ds^2 = f(U)^{-1/2} \left( -dx_0^2 + d\vec{x}^2 \right)
+f(U)^{1/2} d\vec{y}^2\,, \quad f(U) = {{\cal R}^4\over U^4}\,,}
with $\vec{x}=(x_1,x_2,x_3)\,, \vec{y}=(x_4,\ldots x_9)\,, \rho^2
= \vec{y}^2$ and ${\cal
R}^2=\sqrt{4\,\pi\,g_s\,N_c}\,\alpha^{'}\,.$ The induced metric on
the D7-brane embedding is that of $AdS_5\times S^3\,,$ {\sl i.e.}
the D7 fills all the $AdS$ space but is only a contractible cycle
on $S^5$. This breaks the initial $SO(6)$ symmetry on the $S^5$ to
$SO(4)\times SO(2)$. The former corresponds to the isometry group
of the $S^3$ while the latter to the rotations on the $(x_8,x_9)$
plane.

\medskip

The fundamental massive gauge theory is found by separating, a
distance $\ell$, the D7-branes in the $x_8-x_9$ plane,
$x_8^2+x_9^2 = \ell^2$
\eqn\dseven{ds^2=  {U^2 + \ell^2\over  {\cal R}^2} \left( -dx_0^2
+ d\vec{x}^2 \right) + {{\cal R}^2\over  U^2 +
\ell^2}\,d\Omega_4^2\,,\quad d\Omega_4^2=\left( dU^2+ U^2
d\Omega_3^2 \right)\,. }
In flat Minkowsky space the quark mass is proportional to $\ell$.
At the ultraviolet region, $U\gg$, \dseven~ asymptotes to
$AdS_5\times S^3$ providing a conformal invariant gauge theory. In
reducing the energy, decreasing $U$, conformal invariance is lost.

\subsec{Spinning mesons}

We start by determining the possible spectrum for open rotating
strings with ends attached to the D7-branes
\footnote{$\natural$}{The chiral symmetry breaking of the model
was studied in \BabingtonVM.}. The string worldsheet action is
parametrised by $(\tau,\sigma)$ as
\eqn\ac{S=\int d\tau d\sigma {\cal L}\left(X,\partial_\sigma X,
\partial_\tau X\right)\,.}
The string configuration we choose is the same as the one
described in \KruczenskiBE~ but with the insertion of a second
brane probe. It will lie at a fixed point in $S^5$, rotating with
constant angular velocity $\omega$ within a 2-plane in {\sf E}$^3$
and extending in the direction $z={\cal R}^2/U$. The relevant part
of the metric is then
\eqn\rot{ds^2={{\cal R}^2\over z^2} \left(-dt^2 + dr^2 + r^2
d\theta^2 +dz^2\right)\,.}
Defining $z_{m_q}$ as the position of the flavour brane in terms
of the new coordinate $z$, the mass of the quark is
\eqn\quarkm{m_q={{\cal R}^2 \over 2\pi z_{m_q}}\,.}
The rotating string describes a cigar-like surface with the
configuration
\eqn\confcigar{ t=\kappa\, \tau\,,\quad r(\sigma)\,,\quad \theta =
\omega\,\tau\,,\quad z(\sigma)\,.}
Under the latter assumption, \confcigar , the Lagrangian density
takes the form
\eqn\lagdos{{\cal L}= -{{\cal R}^2\over 2\pi\alpha^\prime} {1\over
z^2} \sqrt{\left(\kappa^2-\omega^2 r^2\right) \left( z^{\prime 2}+
r^{\prime 2}\right)}\,,}
where we have already substituted $\omega$ for $\dot{\theta}$ and
denoted $\partial_\sigma$ by primes. The equations of motion for
$z(\sigma)$ and $r(\sigma)$ are equal and they reduce to
\eqn\eqdiff{{1\over z^{'2}+r^{'2}} \left(z' r'' - z'' r'\right)+
\omega^2 {r \over \kappa^2-\omega^2 r^2} z' -2 {r'\over z}=0\,.}
In addition to \eqdiff~ one has to supplement the system with the
proper set of boundary conditions (b.c.). In order to find the
b.c. to be applied we use the minimum action principle: the string
must be stationary under small deformations of the world-line
$$
\delta S= -2 \int d\tau\, d\sigma\, \delta z\, {1\over z^3}
\sqrt{\left(\kappa^2 -\omega^2
r^2\right)\left(r'^{2}+z'^{2}\right)} - \omega^2 \int d\tau\,
d\sigma\, \delta r\, {r\over z^2} \sqrt{{r'^{2}+z'^{2} \over
\kappa^2 -\omega^2 r^2}} $$ \eqn\mes{+\int d\tau\, \delta r\, {r'
\over z^2} \sqrt{{ \kappa^2 -\omega^2 r^2 \over r'^{2}+z'^{2} }}
\Bigg\vert_{\sigma=-\pi/2}^{\sigma=\pi/2}+ \int d\tau\,\delta z
\,{z'\over z^2} \sqrt{{ \kappa^2 -\omega^2 r^2 \over r'^{2}+z'^{2}
}} \Bigg\vert_{\sigma=-\pi/2}^{\sigma=\pi/2} = 0\,. }
Each end of the string lies on one of the D7 branes, {\sl i.e.}
$z$-coordinates obey Dirichlet b.c. While  for \confcigar~ $r$
obeys a Neumann b.c. . {}From \mes~ and in the case of subluminal
velocities this implies that the string ends hit the D7-brane
orthogonally.

The e.o.m. are easily worked out by fixing the gauge

\medskip

\noindent {\it i)} $r(\sigma)=\sigma$
\eqn\eomz{{z''\over 1+ z'^2}- {\omega^2\sigma\over 1-\omega^2
\sigma^2 } \,z' + {2\over z} = 0\,.}
\medskip
\noindent {\it ii)} $z(\sigma)=\sigma$
\eqn\eomrho{{r''\over 1+ r'^2}- 2 {r'\over \sigma } + {\omega^2
\over 1-\omega^2 r^2} \,r = 0\,.}
Notice that \eomz , \eomrho~ are identical to the ones given in
\KruczenskiBE . Our only concert here has to do with the b.c.
imposed on these
\eqn\bcs{ z_{m_q}\,,z_{M_q} = {\rm fixed}\,,\quad \partial_\sigma
z \vert_ {\sigma=\pm {\pi \over 2}} \rightarrow \infty\,.}
In the limiting case, $z_{m_q} = z_{M_q}$ we recover the results
presented in \KruczenskiBE .
\medskip
The system \lagdos~ does not depend explicitly neither on the time
coordinate nor on $\theta$. This implies the existence of two
conserved quantities. Moreover using reparametrisation invariance,
$\kappa=1\,,$ the energy and the spin are given by
\eqn\en{E= {{\cal R}^2 \over 2\pi\alpha'} \int  d\sigma {1\over
z^2} \sqrt{ {z'^{ 2}+r'^{ 2}\over 1 -\omega^2 r^2}}\,,}
\eqn\spin{J={{\cal R}^2 \omega \over 2\pi\alpha'} \int d\sigma
{\left(r\over z\right)^2} \sqrt{ {z'^{ 2}+r'^{ 2}\over 1 -\omega^2
r^2}}\,.}

\medskip

\ifig\ejdos{Chew-Frautschi plot for $m_q=1$ and $M_q=1,2,3$. The
dashed lines represent the limits (2.18) and (2.29) for the equal
and degenerate case. We also have plotted (horizontal thick line)
the energy of a free system of quarks of energy $2m_q\,.$}{
\epsfxsize 2.2 in\epsfbox{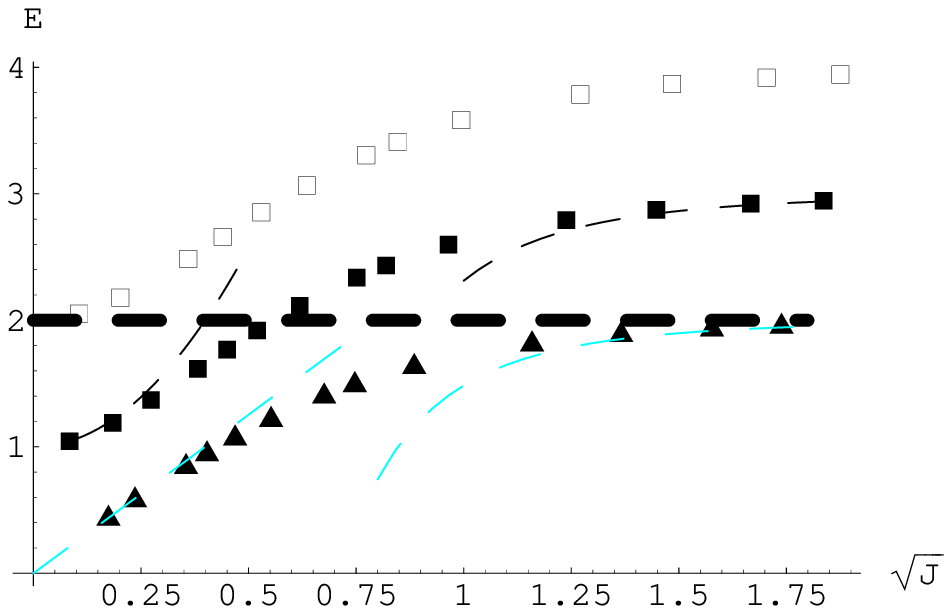}}

We shall study the behaviour of \eomz, \eomrho, \en~ and \spin.
The most straight forward way is a numerical analysis. In \ejdos\
we show the energy versus the squared root of the angular momentum
\footnote{$\flat$}{Even if we plot dots, this by no means is
because any quantisation of the angular momenta. Those intent only
to be representative points.}. As one can see the degenerate
system behaves at low-angular momenta in a completely different
manner as the different masses case. Even though the analysis is
completed it would be illuminating if we recover analytically the
result in two limiting cases. We follow closely \KruczenskiBE~ by
slightly adapting it to the present case.

\bigskip
\noindent {\it 2.1.1. $J\ll \sqrt{g_s N_c}$ Mesons}
\medskip

\ifig\ww{The profile of the string for $J\ll \sqrt{gs Nc}\,$ for
$m_q=1,\, M_q=2$ and $\omega = 4$. The dashed lines represent the
analytic continuation of the string profile. As suggested in
\KruczenskiBE~ those lines correspond to {\sl gluons shells}.
Contrary to the case presented in the above reference in the case
of nondegenerate mass these {\sl gluons shells} are not longer
symmetric. The horizontal lines represent the
probe-branes.}{\epsfxsize 2.2 in\epsfbox{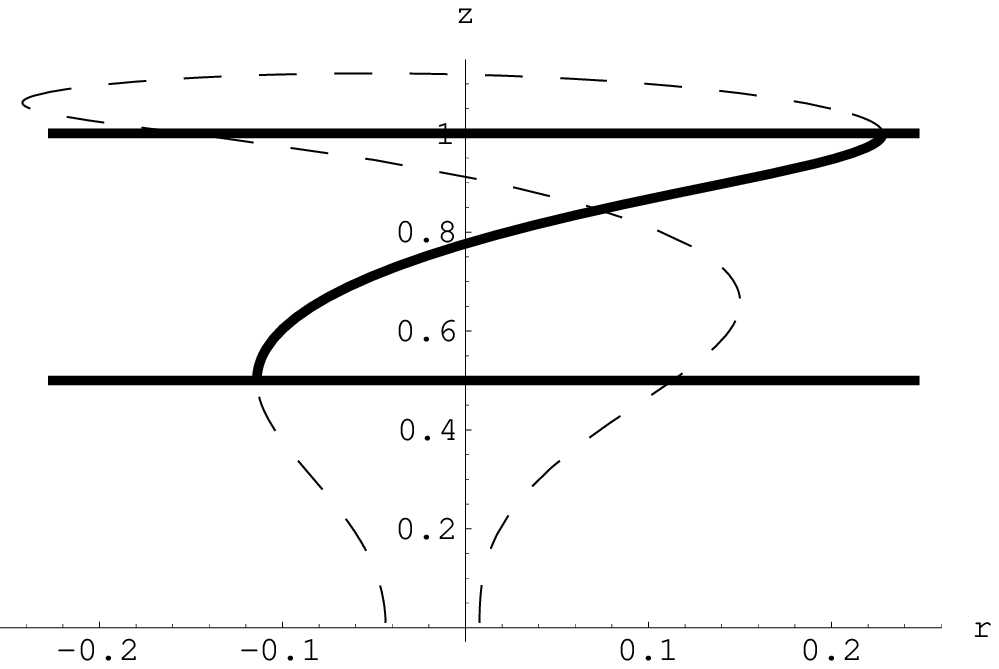}}

Let us analyse the $J\to 0$ limit. As depicted in \ww , the
configuration corresponding to very small $J$ is a string that
directly stretches in the $z$ direction between both flavour
branes, without bending. Obviously, this is qualitatively
different of what happens in the single flavour case. In order to
have $J\to 0$, we need $r \to 0$ (for all the string worldsheet),
as one can realize after a glance at \spin. Therefore, we are in
the limit corresponding to rotating short strings. Neglecting
higher order terms in $r$, \eomrho~ becomes
\eqn\eomshort{r''- 2 {r'\over z } + \omega^2  \,r = 0\,,}
with $r(z_{m_q})=\epsilon$~ and where only the leading order in
epsilon is taken into account. Supplementing this equation with
the boundary conditions $r'(z_{m_q})=0,\ r(z_{m_q})=\epsilon\ll$,
we find
\eqn\rhoshort{r(z)= {\epsilon  \over  z_{m_q}\,\omega} \left[z
\omega \cos\left((z_{m_q}-z)\omega\right)
+\sin\left((z_{m_q}-z)\omega\right)\right]\,.}
Moreover, by fixing $r'(z_{M_q})=0$
\eqn\wshort{\omega={\pi \over z_{m_q}-z_{M_q}}\,.}
This is a limiting upper value of the angular velocity,
contrariwise to the one-mass case, where $\omega \to \infty$ when
$J\to 0$.

Notice that $m_q\,r(z_{m_q})=M_q\,r(z_{M_q})\,.$ This is the
momentum conservation relation for two rotating masses, {\sl i.e.}
the string does not carry any energy in this limit acting only as
a geometrical constraint.

The energy-spin relation is given in this case by
 \eqn\reggeshort{E=M_q - m_q +
{2 \pi^2 \alpha'\over {\cal R}^2} {m_q M_q\over M_q+m_q}\ J\,.}
This linear behaviour relating the energy and the angular momentum
at high energy has no field theory interpretation.

\bigskip
\noindent {\it 2.1.2. $J\gg \sqrt{g_s N_c}$ Mesons}
\medskip

\ifig\wilplot{The $z(r)$ profile for the degenerate, l.h.s., and
nonequal mass case, r.h.s.$\,.$ The dashed lines represent the
analytic continuation of the profile following eqs. \eomz,
\eomrho. The horizontal solid lines represent the two flavour
branes. $z=0$ represents the AdS boundary.}{\epsfxsize 2.2
in\epsfbox{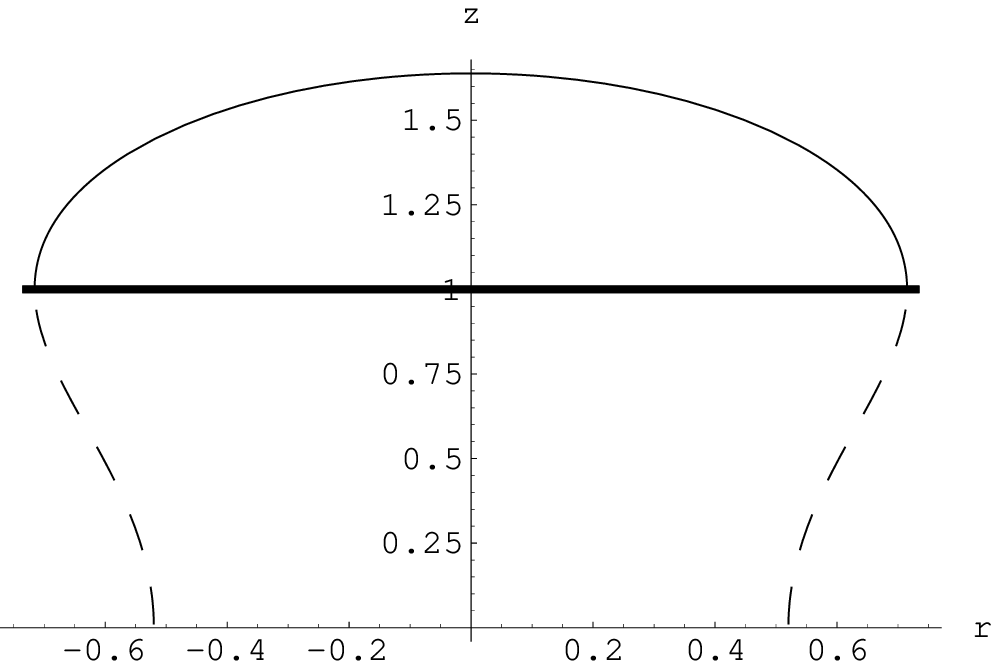} \epsfxsize 2.2 in\epsfbox{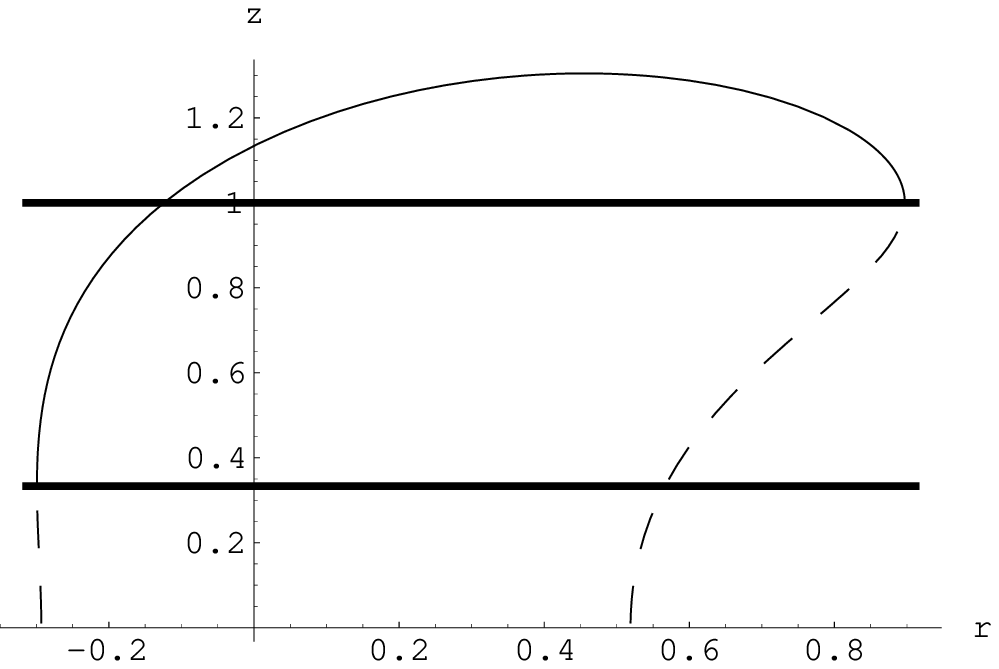}}

The other possible limit where \en~ and \spin~ can be tackled
analytically is $\omega \to 0$ which corresponds to rotating long
strings that give large $J$ mesons. We shall generalize the
analysis of \KruczenskiBE~ in order to incorporate the effect of
having two different quark masses. For small angular velocity, the
profile of the string must be similar to that of a static string
hanging between both branes. This is given by
\eqn\static{r_{{\rm st}}(z) = r_0 + \int_z^{z_0} dx {x^2\over
\sqrt{z_0^4 - x^4}}\,,}
where $z_0 \to \infty$ in the limit we are considering and it
refers to the extremal point. We have defined $r_0$ as the $r$
coordinate of the point where the string bends back. It cannot be
set to zero as $r=0$ is by definition the center of rotation of
the string and both points do not coincide for nondegenerate
masses (see the asymmetric plot in \wilplot). Let us point out
that the distance between the string edges is
\refs{\ReyIK,\MaldacenaIM}
\eqn\totlength{r_{m_q}+r_{M_q}=2 \,{\cal C}\, z_0\,,}
with
 \eqn\c{{\cal C}= \int_1^\infty {dy\over y^2\sqrt{y^4-1}} =
{\sqrt{\pi}\Gamma\left(3/4\right)\over\Gamma\left(1/4\right)}\,.}
By considering $r(z)=r_{{\rm st}}(z)+\delta r(z)$ and keeping the
leading order in $\delta r(z)$, $\omega^2$, \eomrho~ reads
\eqn\largesusyeq{\delta r''-{2 \over z}(1+3r_{{\rm st}}'^2)\delta
r'= -\omega^2 r_{{\rm st}} {1+r_{{\rm st}}'^2 \over 1-\omega^2
r_{{\rm st}}^2}\,\,.}
This equation can be straightforwardly integrated
 \eqn\soldifeq{\delta
r'(z)= {z^2 \over (z_0^4 - z^4)^{3 \over 2}} \left[ {\rm const} +
\omega^2 z_0^4 \int_{z_0}^z {\sqrt{z_0^4 - x^4 } \over x^2}
\,{r_{{\rm st}}(x) \over 1-\omega^2 r^2_{\rm st}(x)}dx\right]\,, }
and in addition has to be supplemented with the boundary
conditions \bcs. They read
\eqn\bcslong{ r_{{\rm st}}'(z_{m_q})+\delta r'(z_{m_q})= r_{{\rm
st}}'(z_{M_q})+\delta r'(z_{M_q})=0\,.}
As we are in the long string regime, one can approximately think
on the profile of the string in the following way: It stretches
vertically from $z_{M_q}$ to $z_0$ at constant $r_{M_q}$, then
stretches horizontally from $r_{M_q}$ to $r_{m_q}$ at constant
$z_0$ and finally stretches again vertically from $z_0$ to
$z_{m_q}$ at constant $r_{m_q}$. A glance at \soldifeq~ reveals
that the integral is dominated by the vertical regions {\it i.e.}
$x\ll z_0$. Moreover, the constant can be set to zero as the point
with $\delta r' =0$ is in the horizontal region, at least when the
quark masses are of the same order of magnitude (this reasoning
fails when ${M_q \over m_q} \to \infty$, because when there is an
infinite mass $r_{{\rm st}}'(z_{M_q})=0$ so $\delta
r'(z_{M_q})=0$, {\it i.e.} the $\delta r' =0$ point is in the
vertical region).

Then, one finds from \bcslong
 \eqn\susyomega{\omega^2 = {z_{m_q}
\over r_{m_q} z_0^2}= {z_{M_q} \over r_{M_q} z_0^2}\,\,,}
where we have used the relation $z_0 \gg z_{m_q},z_{M_q}$ and, as
can be proved from \totlength,\susyomega, $\omega^2
r_{M_q}^2,\omega^2 r_{m_q}^2 \ll 1$. {}From \susyomega~ it is
immediate that
 \eqn\cdmsusy{m_q r_{m_q}=M_q r_{M_q}\,.}
This is the momentum conservation relation for two rotating
nonrelativistic masses, and signals that the horizontal part of
the string carries a neglectable amount of energy.

Using the relations above, it is easy to compute the angular
momentum from \spin~ to leading order
 \eqn\sp{J\approx  {{\cal
R}^2 {\cal C}\over \pi \alpha^{'} z_0 \,\omega}+\ldots\,.}
The energy \en, considering the leading and subleading orders is
\eqn\e{E \approx {{\cal R}^2 \over 2 \pi \alpha^{'}} \left(
{1\over z_{m_q}} + {1\over z_{M_q}}- {{\cal C} \over z_0} \right)
+\ldots \,.}
Using \totlength~, \susyomega~ and \sp~ this expression can be
recast as
\eqn\nonrela{E= M_q + m_q - E_{\rm binding}\,,\quad E_{\rm
binding}= {2m_q M_q \over (m_q+M_q)}\, {\kappa^4\over 4 J^2}\,,}
where
\eqn\kappaeq{\kappa^4=\left({{\cal R}^2 \over 2 \pi
\alpha'}\right)^2\, 16\,{\cal C}^4={16\over \pi} g_s N_c \,{\cal
C}^4\,.}
We deal with a system of two nonrelativistic particles bounded in
a Coulombic potential, $V(r) \sim -\kappa^2/r\,,$ where only a
finite energy cost is needed to separate the two masses. Notice
that the binding energy is proportional to the reduced mass of the
system. At relatively large separations, $J\gg\,,$ the hadron mass
asymptotes to $M_q+m_q\,.$ While for given masses and strong
coupling $E_{\rm binding} \ll (M_q+m_q)$, hence been the string
stable. When the limit $M_q \rightarrow \infty$ with $ m_q$ held
fixed is taken the binding energy must become independent of the
coupling \KarchXE\ and our procedure fails.

\subsec{Quark-antiquark energy}

The Wilson loop provides an order parameter for pure gluodynamics.
Even if in the presence of dynamical quarks this feature is washed
out, it still shows a strong sensitivity in the transition region
and therefore is widely used to describe crossover in full QCD. A
possible gravity analog for infinitely heavy-quarks was presented
in \refs{\ReyIK,\MaldacenaIM}.

The addition of dynamical flavour test sources \KarchSH\ ,
D7-branes, allows for the inclusion of light-quark masses
\KruczenskiUQ . It is of interest to investigate the behavior of
the $\overline{q}q$ energy as a function of their relative
separation in this case. One of the expected new features is the
screening in the potential due to light-quarks. To compute the
$q\bar{q}$ energy we locate an open string with the two ends on
the probe-branes and stretching from them to the origin in the $U$
coordinate. At some point $U_0$, before hitting the horizon, it
bends back. The relevant part of the metric reduces then to
\eqn\wilmetric{ds^2= \alpha^\prime\left[ {U^2\over {\cal R}^2}
\left(-dt^2+dr^2\right)+ {{\cal R}^2\over U^2} dU^2\right]\,. }
The string configuration is  parametrised as
$$
U=U(\sigma)\,,\quad r=\sigma\,,\quad t=\tau\,.$$ This leads to the
lagrangian density
\eqn\lagden{{\cal L}={1\over 2\pi\alpha'} \int_{r_{\rm min}}^{r_{\rm
max}} dr \sqrt{U^{\prime 2} + {U^4\over {\cal R}^4 }}\,.}
The independence of \lagden~ w.r.t. the world-sheet variable
$\sigma$ implies the existence of a conserved quantity
\eqn\conservation{{U^4\over {\sqrt{U^{\prime 2} +{U^4\over {\cal
R}^4} }}}={\cal R}^2 U_0^2\,.}
\ifig\ddd{The profile of the string in the $U\,-\,r$ plane for
different lengths. The figure on the left corresponds to long
strings  and the one on the right to short strings. In both cases
we take $m_q=1,\, M_q=2$}{\epsfxsize 2.2 in\epsfbox{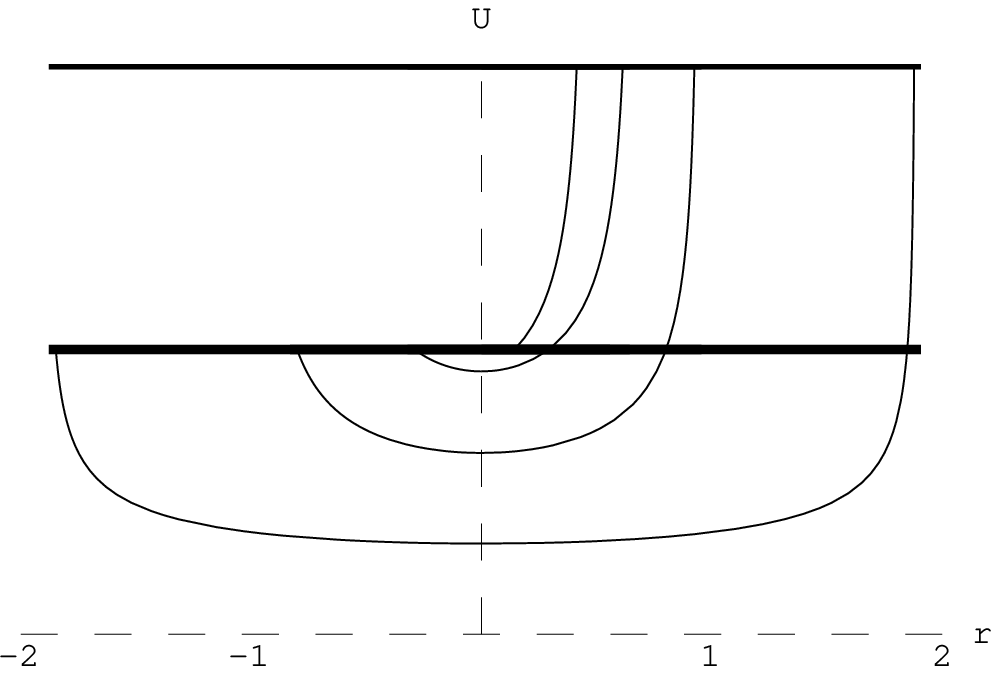}
\epsfxsize 2.2 in\epsfbox{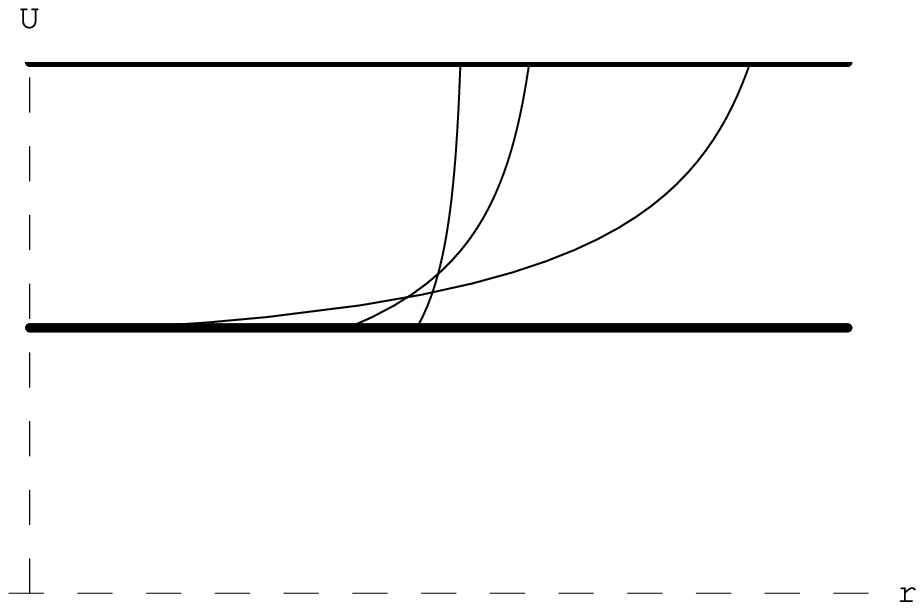}}
\noindent In \ddd~ we have plotted the profile of the string
obtained by numerically solving \conservation. {}For long strings
there is a turning point while short strings stretch directly
between probe branes. Notice that the long string limit is
achieved by approaching $U_0\rightarrow 0\,.$

The separation between the two ends of the string can be easily
computed from \conservation
\eqn\lenght{{\rm L}=2 {{\cal R}^2 \over U_0} \int_1^{2\pi
{m_q\over U_0}} {dy\over y^2\sqrt{ y^4 -1 }} + {{\cal R}^2 \over
U_0}\int_{2\pi {m_q\over U_0}}^{2\pi {M_q\over U_0} }{dy\over
y^2\sqrt{y^4-1}} \,.}
In the latter expression the second term in the r.h.s. is due to
the breaking of the mass degeneracy. With the use of \lenght~ the
energy \conservation~ can be casted as
\eqn\en{E= {U_0\over \pi} \int_1^{2\pi {m_q\over U_0}} dy
{y^2\over \sqrt{y^4-1}}+{U_0\over 2\pi} \int_{2\pi{m_q\over
U_0}}^{2\pi{M_q\over U_0}} dy {y^2\over \sqrt{y^4-1}} \,.}
However, notice that \lenght~ can only describe configurations for
which the string is longer than
 \eqn\shlength{ L_{lim} =
{{\cal R}^2 \over 2 \pi m_q } \int_1^{{M_q \over m_q}} {dy\over
y^2\sqrt{y^4-1}} \,.}
This value obtained for $U_0= 2\pi m_q$. For $L<L_{lim}$, the
string does never reach values of $U$ smaller than $2 \pi m_q$, as
can be seen in \ddd . Therefore, the separation between the string
ends and the energy are given simply by
 \eqn\short{{\rm L}={{\cal R}^2 \over
U_0}\int_{2\pi {m_q\over U_0}}^{2\pi {M_q\over U_0} }{dy\over
y^2\sqrt{y^4-1}} \,, \,\,\,\,\,\,\, E= {U_0\over 2\pi}
\int_{2\pi{m_q\over U_0}}^{2\pi{M_q\over U_0}} dy {y^2\over
\sqrt{y^4-1}} \,.}

\ifig\ELplot{The energy in terms of the length for $m_q=1$ and,
from bottom to top,  $M_q= 1,2,3$, l.h.s.{} For the $M_q=2$ case
we show the asymptotic (2.41) and (2.42). In the right panel we
show the case of infinite heavy mass for {\sl i)} the D3-D7 system
(2.38), thin curve and {\sl ii)} for the D4-D6 system (3.29) thick
curve.}{\epsfxsize 2.2 in\epsfbox{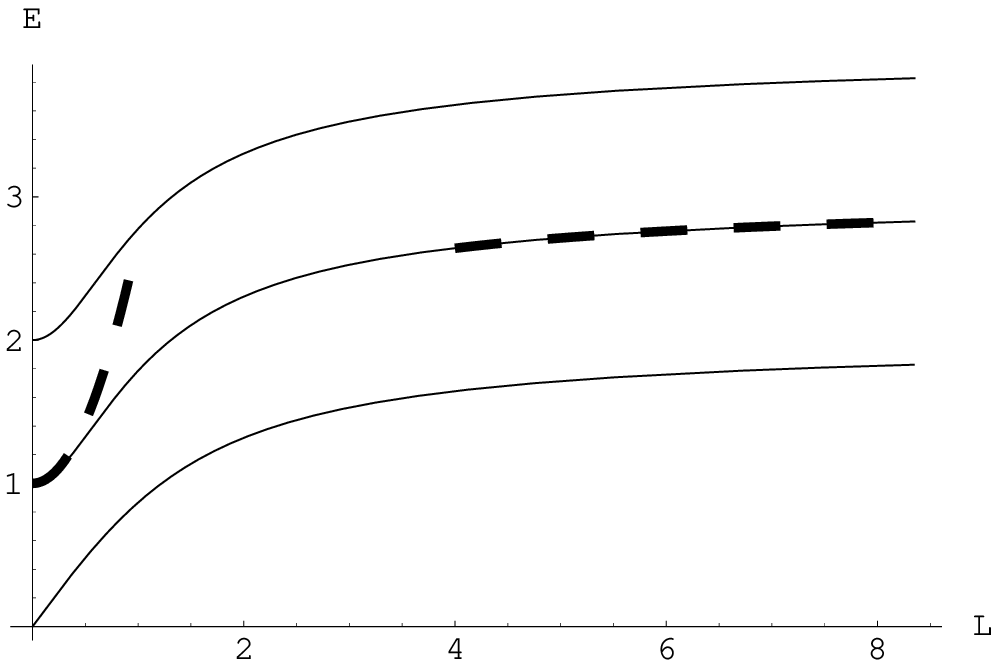} \epsfxsize 2.2
in\epsfbox{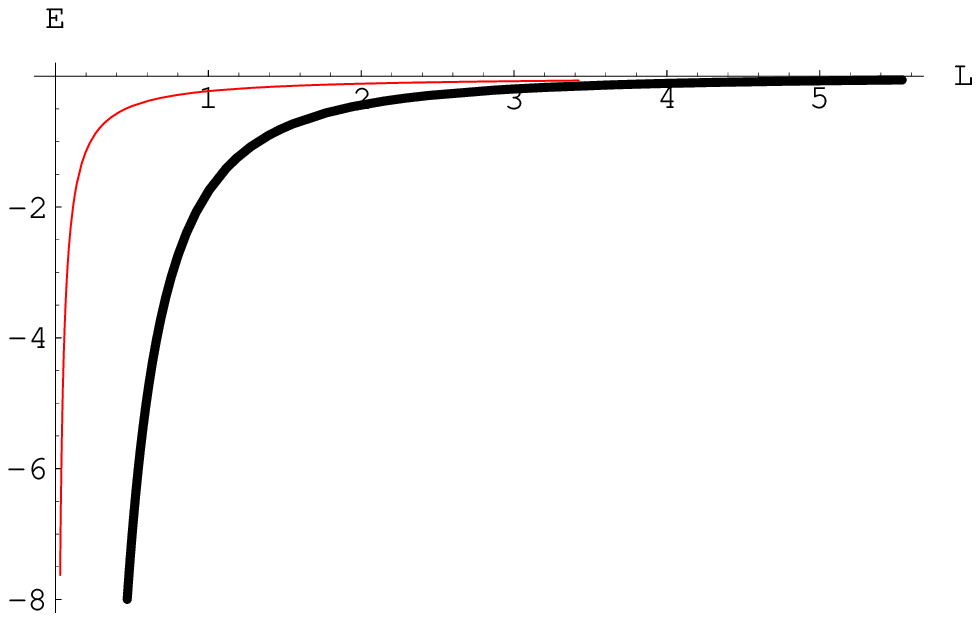} }

In \ELplot , we have plotted the energy as a function of the
distance between the string ends. For large $L$, the effect of
incorporating a heavier mass shifts the curves upwards but without
modifying its characteristics. But for small $L$ a novel effect
w.r.t. the equal mass case appears: instead of a pure linear
behaviour there is a slight positive curvature, {\sl a kind of
parabola}, that increases with the mass difference. Notice that in
this case the slope, as approaching $L\to 0^+$, decreases
indicating a smaller running of the $\beta$-function.

\bigskip
\noindent {\it 2.2.1. Mass Subtraction}
\medskip

Notice that \en\ is not a renormalised expression, and therefore
it will diverge as soon as the upper cutoffs are moved to the
boundary at infinity. In that case, we shall deal with an
infinitely heavy, degenerate pair of quarks. To render \en\ finite
in this case, we subtract the mass of the static quarks given by a
couple of strings, one for each quark, stretching from the
boundary to the horizon along the transverse coordinate $U(\sigma)
= \sigma\,, \quad X_0 = \tau$~\MaldacenaIM. With this \en\ reads
\eqn\energiabis{E_{\rm ren} = {U_0\over \pi}\left[
\int_1^{\infty}\!dy\,\left( {y^2\over \sqrt{y^4-1}}-1\right) - 1
\right]\,. }
\smallskip
Different shapes in the string providing the subtraction must
correspond to different renormalisation schemes in field theory.
We have checked, in the degenerate case, that \en~ and not the
analog of \energiabis~ for finite masses gives the minimum energy
for the string configuration: we locate a pair of quarks,
$q\bar{q}$, on the D7-brane with a fixed distance among them,
$L_0$. The dynamics is purely dictated by the motion of the
$q\bar{q}$ system under the potential created by the stack of the
D3-branes at the origin. Minimizing  \en~ and \energiabis~ as a
function of $m_q$, $E(L_0,m_q)$, one can verify that the energy of
the system is minimum when \en\ is used. In increasing $m_q$ to
high values the difference between \en~ and \energiabis~ is
reduced until we reach the limit $m_q\rightarrow \infty$ where the
role is interchanged and the energy minimum is given by
\energiabis.

\bigskip

It is instructive to discuss some approximation to different
scales:

\bigskip
\noindent {\it 2.2.2. Long strings}
\medskip

At low-energy we shall take the following set of assumptions
$$
{\rm Length}=L\, {\rm fixed~ and~ large}\,; \quad m_q\,,M_q \gg 2
\pi \,.
$$

Solving iteratively (always at leading order) one can find the
turning point for a given string length
\eqn\lengthone{U_0={{\cal R}^2\over L}{\cal C} \left[ 2 - {1\over
3} {{\cal C}^2 {\cal R}^6\over \left( \pi L\right)^3
}\left({1\over m_q^3} + {1\over M_q^3}\right)\right]\,,}
with ${\cal C}$ given in \c . {F}or $M_q\rightarrow m_q$~ we
recover the symmetric case \KruczenskiBE. Corrections due to heavy
masses are suppressed.

Within the same approximation the energy gives
$$
 E=M_q+ m_q + {U_0\over \pi} \left[ \int_1^\infty dy \left(
{y^2\over \sqrt{y^4-1}} -1\right) -1 \right] -{U_0\over 2\pi}
\int_{2\pi {m_q\over U_0}}^\infty dy \left({y^2\over \sqrt{y^4-1}}
-1\right)$$ \eqn\ee{  - {U_0\over 2\pi} \int_{2\pi {M_q\over
U_0}}^\infty dy \left({y^2\over \sqrt{y^4-1}} -1\right)\,.}

Without performing the limit $M_q,m_q \rightarrow \infty$~ there
is is no need of bare mass subtraction as in
\refs{\ReyIK,\MaldacenaIM} (see discussion above). Using the
leading order in \lengthone~ we obtain
\eqn\enlargetwo{E\approx M_q+m_q - {\kappa^2\over L}\left[ 1 -
{1\over 12} {{\cal R}^6 {\cal C}^2\over (\pi L)^3} \left( {1\over
m_q^3}+{1\over M_q^3}\right)\right]\,.}
Notice that the the binding energy between quarks is attractive
and corresponds at leading order to a system of two
nonrelativistic particles.


\bigskip
\noindent {\it 2.2.3. Short strings}
\medskip

We shall consider now the case where the string is short. The
limit $L\to 0$ is clearly achieved by taking $U_0\to 0$, \short.
Then, it is easy to prove that only keeping the first nontrivial
contribution of the Taylor expansion of the energy in terms of the
string ends separation, one has a quadratic behaviour
\eqn\enshort{ E=M_q - m_q + {6 \pi^2 \over  {\cal R}^4 } \left(
{1\over m_q^3}-{1\over M_q^3}\right)^{-1}L^2\,,}
contrariwise to any field theory expectation.

\lref\BrisudovaUT{ M.~M.~Brisudova, L.~Burakovsky and T.~Goldman,
``Effective functional form of Regge trajectories,'' Phys.\ Rev.\
D {\bf 61}, 054013 (2000) [arXiv:hep-ph/9906293].
}
%

\lref\GeorgiUM{ H.~Georgi, ``An Effective Field Theory For Heavy
Quarks At Low-Energies,'' Phys.\ Lett.\ B {\bf 240}, 447 (1990).
}

\lref\IsgurVQ{ N.~Isgur and M.~B.~Wise, ``Weak Decays Of Heavy
Mesons In The Static Quark Approximation,'' Phys.\ Lett.\ B {\bf
232}, 113 (1989).
}


\lref\WittenZW{ E.~Witten, ``Anti-de Sitter space, thermal phase
transition, and confinement in  gauge theories,'' Adv.\ Theor.\
Math.\ Phys.\  {\bf 2}, 505 (1998) [arXiv:hep-th/9803131].
}

\lref\PonsDK{ J.~M.~Pons, J.~G.~Russo and P.~Talavera,
``Semiclassical string spectrum in a string model dual to large N
QCD,'' Nucl.\ Phys.\ B {\bf 700}, 71 (2004)
[arXiv:hep-th/0406266].
}

\newsec{QCD$_{3+1}$: D4-D6 brane model}

A more realistic model, in the sense of being four-dimensional and
not supersymmetric, can be obtained by placing a stack of $N_c$
$D4$ branes at the origin and wrapping them on a $S^1$ cycle. The
latter are the source for fermion mass at one loop once
antiperiodic b.c. are assumed \WittenZW. As previously we can set
an $N_f$ $D6$-probe branes displaced a distance in some subspace
orthogonal to both brane sets.

\medskip
\parskip=0pt\parindent=30pt\noindent
The field components at leading order in $\alpha'$~are

\item{{\sl i)}} The $10d$ background metric
$$
ds^2_{10} = \left({U\over R}\right)^{3/2} \left(-dx_0^2+dx_1^2
+...+dx_3^2+ {4\, R^3\over 9 \, U_h}\,f(U)\, d\theta_2^2 \right)
$$
\eqn\metricM{ + \left({R\over U}\right)^{3/2}  {dU^2\over f(U)} +
R^{3/2}\, U^{1/2}\, d\Omega_4^2\,,}
with $f(U)= 1-{U_h^3\over U^3}\,, R=\left(\pi N_c g_s\right)^{1/3}
\alpha'^{1/2}\,$~and $U_h$ denoting the horizon location.

\item{{\sl ii)}} A constant four-form field strength (in string
units)
\eqn\w{F_4 = {3\over 8 \pi^3 g_s}\, R^3 \omega_4\,,}
with $\omega_4$ been the volume form of the transverse $S^4$.

\item{{\sl iii)}} Finally, the dilaton field
\eqn\dil{ e^{\Phi}=g_s \left({U\over R}\right)^{3/4}\,.}
\medskip
The relation with the gauge field theory parameters is given by
$$
2 \pi \lambda = {3 R^{3/2} U_h^{1/2}
\over \alpha'}= g^2_{\rm YM}\, N_c \, .
$$
\medskip

To be in the supergravity approximation the following requirements
of the fields have to be fulfilled

\item{{\sl i)}} The smallness of the scalar field \dil. This
restricts the reliability of the background to the infrared
region.

\item{{\sl ii)}} The smallness of the curvature invariant and
higher order derivatives.
\bigskip

This two restrictions amounts to take the 't Hooft limit $g_{\rm
YM}^2 N_c \gg 1$.
\medskip

In addition, in order to describe QCD$_{3+1}$ one has to demand
\item{{\sl iii)}} that the energy scale is such that its
associated transverse coordinate $U$ in the supergravity
description corresponds to values where the background is indeed
dual to QCD$_{3+1}$ rather than QCD$_{4+1}$.
\medskip
The third restriction can be examined from the perspective of the
quark-antiquark energy, which will be considered in section (3.3).
In fact, it is only in the regime of large distances $L$ between
the quark and the antiquark that we reproduce the area law
describing the feature of confinement of QCD. At small distances
the $S^1$ opens up and the background no longer describes
QCD$_{3+1}$ but QCD$_{4+1}$. If the position of the $D6$ brane,
where the quark-antiquark stay, is far away in the transverse
coordinate, we shall deal with a Coulomb phase for the $4+1$ field
theory. Large distances means that the string of the Wilson loop
goes deep into the bulk along the transverse coordinate until it
bends in a minimum near the horizon. The bending must occur at
some $U_0$ such that ${U_0-U_h\over U_h} \ll 1$. This is a third
condition that must be met in order to stay in the region where
the background exhibits the area law behaviour for the Wilson
loop. Note that this condition will guarantee that $U-U_h \ll
M_{kk}$, which is the regime where the theory is effectively four
dimensional \PonsDK.

Finally, let us point out that the mass of the quark is given by
the energy of a string stretching from the horizon to a flavour
brane
\eqn\qmass { m_q={1\over 2\pi\alpha'} \int_{U_h}^{U_{D6}} {1 \over
\sqrt{1- \left({U_h^3 \over U^3}\right)}} dU\,\,.}

\subsec{Spinning mesons}
\lref\KruczenskiME{ M.~Kruczenski, L.~A.~P.~Zayas, J.~Sonnenschein
and D.~Vaman, ``Regge trajectories for mesons in the holographic
dual of large-N(c) QCD,'' arXiv:hep-th/0410035.
}

\lref\ida{M.~Ida, ``Relativistic motion of massive quarks
joined by a massless string", {\sl Progr. Theor. Phys.}
{\bf 59}, 1661 (1978).}

\lref\burakovsky {L.~ Burakovsky, ``String model for analytic nonlinear
Regge trajectories", hep-ph/9904322.}

\lref\loopref {F.~ Bigazzi, A.~ Cotrone, L.~  Martucci and L.~ Pando
  Zayas, ``Wilson loop, Regge trajectory and hadron masses in a Yang-
    Mills theory from semiclassical strings", hep-th/0409205.}

\lref\LyubimovKM{ V.~A.~Lyubimov, ``Backward Scattering Of Pions
On Nucleons,'' Sov.\ Phys.\ Usp.\  {\bf 20}, 691 (1977) [Usp.\
Fiz.\ Nauk {\bf 123}, 3 (1977)].
}
\lref\BrandtGI{ A.~Brandt {\it et al.}  [UA8 Collaboration],
``Measurements of single diffraction at s**(1/2) = 630-GeV:
Evidence for  a non-linear alpha(t) of the pomeron,'' Nucl.\
Phys.\ B {\bf 514}, 3 (1998) [arXiv:hep-ex/9710004].
}

We turn once more to the analysis of rotating strings.
The initial point is \ac\ and the string configuration is
identical to \confcigar\ . The relevant part of the metric is
\eqn\mdquatredsis{ ds^2= \left({U\over R}\right)^{3/2} \left(-dt^2
+ dr^2 + r^2 d\theta^2\right) +  \left( {R\over U}\right)^{3/2}
{dU^2 \over f(U)} \,.}
That leads to the action \KruczenskiME\
\eqn\acquatredsis{ S= {1\over 2\pi \alpha'} \int d\sigma d\tau
\sqrt{ U^{3} \left( \kappa^2 -\omega^2 r^2\right) \left(
{U'^{2}\over U^{3} f(U)} + {r'^{2}\over R^3}\right)}\,.}
As in the D3-D7 case the variation of the action imposes that the
strings end orthogonally to the D6-branes.

In general the e.o.m derived from \acquatredsis~ are difficult, if
not impossible, to be tackled in an analytic way, but as
previously the solutions can simplify by choosing the proper gauge
fixing conditions ($\kappa=1)$.

\medskip

\noindent {\it i)} $U(\sigma)=\sigma$
\eqn\eomr{ r''- {3 \over 2 R^3 \sigma }\left(U_h^3-\sigma^3\right)
\,r'^{3} + {\omega^2\over 1-\omega^2 r^2} \,r\, r'^{2} + {3\over
2\sigma}{ \left(U_h^3-2\sigma^3\right)\over
\left(U_h^3-\sigma^3\right)} \,r'- {R^3 \omega^2\over
\left(U_h^3-\sigma^3\right) \left(1-\omega^2 r^2\right)}\,r= 0\,.}
\medskip
\noindent {\it ii)} $r(\sigma)=\sigma$
$$
U''+{R^3 \omega^2 \sigma \over \left(1-\omega^2
\sigma^2\right)\left(U_h^3-U^3\right)} \,U'^3 - {3\over2 U}
\left({U_h^3-2U^3\over U_h^3-U^3}\right) U'^2-{\omega^2\sigma
\over 1-\omega^2\sigma^2}U'$$ \eqn\eomrho{ -{3\over
2R^3}\left({2U_h^3-U^3\over U_h^3-U^3}\right) U^2 + {3U_h^6\over 2
R^3 U\left(U_h^3-U^3\right)}= 0\,.}
In addition we impose the b.c. $ U_{D6}^{(i)} = {\rm fixed}\,,
\partial_\sigma U_{D6}^{(i)} \vert_ {\sigma=\pm {\pi \over 2}} \rightarrow
\infty\,, (i=1,2)\,.$

The invariance of \acquatredsis~ with respect to $t$~ and $\theta$
signals the energy and spin conservation
\eqn\enqcd{E={1\over 2\pi\alpha'}\int d\sigma \left(U\over
R\right)^{3/2}  \sqrt{ {r'^2+ {R^3 U'^2\over U^3-U_h^3}\over
1-(\omega r)^2}}\,,}
\eqn\sqcd{J={\omega\over 2\pi\alpha'}\int d\sigma\,r^2\left(U\over
R\right)^{3/2} \sqrt{ {r'^2+ {R^3 U'^2\over U^3-U_h^3}\over
1-(\omega r)^2}}\,.}
{}For the rest of this section, we set $2 \pi \alpha'=1\,.$

\bigskip
\noindent {\it 3.1.1. Meson spectrum at large $J$}
\medskip

Once again, the regime of large angular momentum corresponds to
long strings rotating with small angular velocity $\omega \to
0\,.$

\ifig\profiles{The profile for long strings, $\omega \to 0 \,.$
Equal mass case and different masses (l.h.s. and r.h.s.
respectively). The dotted line denotes the horizon position, the
horizontal full lines the flavour brane location and the dashed
line the analytic string continuation.}{\epsfxsize
2.2in\epsfbox{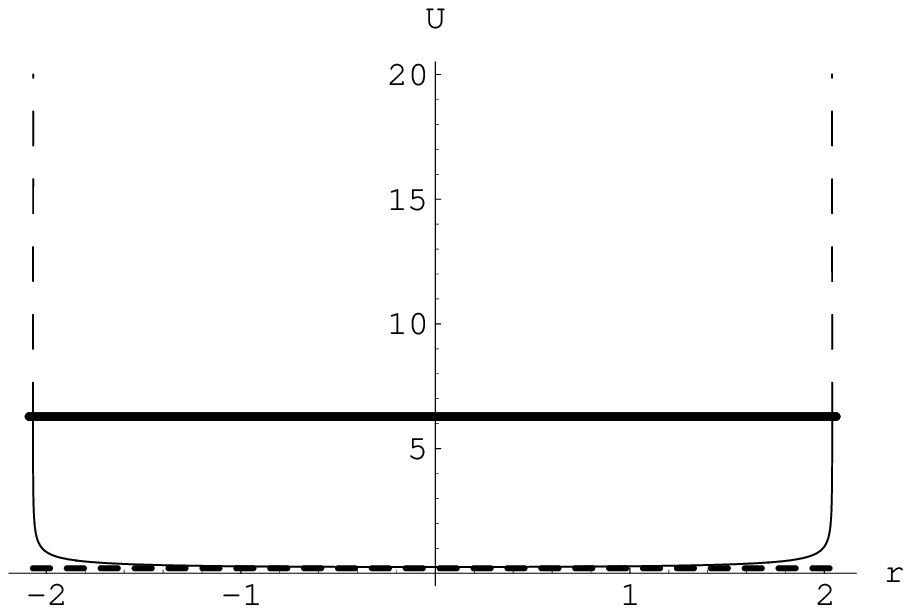}\epsfxsize 2.2
in\epsfbox{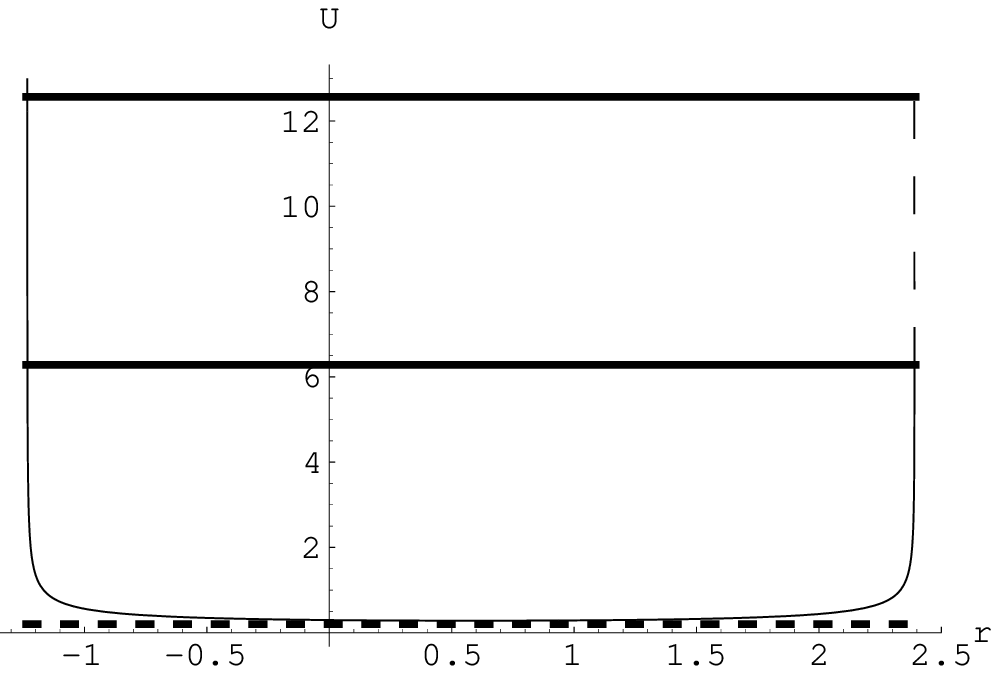} }
In \profiles~ we show a numerical solution for the string profile.
This is approximately given by that of a static long string
\eqn\lenden{ \partial_U r = \left(R U_0\right)^{3/2} {1\over
\sqrt{\left(U^3-U_0^3\right)\left(U^3-U_h^3\right) }}\,,}
which is characterised by {\sl almost} straight strings. Notice
that \lenden~ diverges in the limit $U_0\rightarrow U_h$. As
previously $U_0$ denotes the string turning point.

The numerical solutions to \eomr, \eomrho~ are almost identical to
those depicted in \ww~and \wilplot. Even possessing these
numerical solution is not too much illuminating. As first noticed
in \KruczenskiME, an inspection of \profiles~ reveals that for
slowly rotating string, $J\gg \sqrt{g_s N_c}$, a good
approximation is to consider three parts of the string: two
stretching in the $U$ direction from $U_{D6}^{(1)}$ and
$U_{D6}^{(2)}$ to $U_h$ and the third horizontal piece where the
string stretches in the field theory directions ($r$) at $U=U_h$.
This allows to split approximately the integrals \enqcd, \sqcd~ in
three regions. The two vertical regions, correspond to $r'^2 \ll
{R^3 U'^2\over U^3-U_h^3}$~ and the horizontal region ($U\approx
U_h$)~ to $r'^2 \gg {R^3 U'^2\over U^3-U_h^3}$\footnote*{In order
to see this, consider $\sigma = U$, so both terms go to infinity
but the first one dominates as the string profile at small
$\omega$~ must be similar to the static one, so \lenden \ holds
approximately.}. Therefore, the energy reads
$$
E=$$ $$ \left({U_h\over R}\right)^{3 \over 2}
\int_{-r_{m_q}}^{r_{M_q}}\!\!{ dr \over \sqrt {1-\omega^2 r^2}}+
{1 \over \sqrt {1-\omega^2 r_{m_q}^2}}\int_{U_h}^{U_q}\!
{U^{3\over 2} dU \over \sqrt {U^3-U_h^3}}+ {1 \over \sqrt
{1-\omega^2 r_{M_q}^2}}\int_{U_h}^{U_Q} {U^{3\over 2} dU \over
\sqrt {U^3-U_h^3}}$$ \eqn\Eapprox{=\left({U_h \over R}\right)^{3
\over 2} \omega^{-1}\left(\arcsin (\omega r_{m_q}) +\arcsin
(\omega r_{M_q})\right) +{m_q \over \sqrt {1-\omega^2 r_{m_q}^2}}+
{M_q \over \sqrt {1-\omega^2 r_{M_q}^2}}\,,}
whereas the angular momentum can be expressed as
$$
J=$$ $$ \left({U_h\over R}\right)^{3\over 2}
\int_{-r_{m_q}}^{r_{M_q}} \!{\omega\,r^2\,dr \over 2 \sqrt
{1-\omega^2 r^2}} + {\omega \ r_{m_q}^2 \over \sqrt {1-\omega^2
r_{m_q}^2}}\int_{U_h}^{U_q} \!{U^{3\over 2} dU \over \sqrt
{U^3-U_h^3}}+ {\omega\ r_{M_q}^2 \over \sqrt {1-\omega^2
r_{M_q}^2}}\int_{U_h}^{U_Q} \!{U^{3\over 2} dU \over \sqrt
{U^3-U_h^3}}$$
$$
=\left({U_h \over R}\right)^{3 \over 2}\omega^{-2} \left(\arcsin
(\omega r_{m_q}) - \omega r_{m_q} \sqrt{1-\omega^2 r_{m_q}^2}+
\arcsin (\omega r_{M_q}) - \omega r_{M_q} \sqrt{1-\omega^2
r_{M_q}^2}\right)+
  $$
\eqn\Japprox{+{\omega r_{m_q} m_q \over \sqrt {1-\omega^2
r_{m_q}^2}}+ {\omega r_{M_q} M_q \over \sqrt {1-\omega^2
r_M^2}}\,.}
Notice that this is just a straightforward generalization of the
results presented in \KruczenskiME. Hereafter $r_{m_q}$ and
$r_{M_q}$ stand for the radius of rotation of the light and heavy
quark respectively.

Recall that not all of the parameters in the above expressions are
independent. For given values of the masses, once $\omega$~ is
fixed, the values of $r_{m_q}$, $r_{M_q}$~ follow from it as a
consequence of the boundary conditions that must be imposed on the
string. In order to compute this relation, we will follow the
steps of \KruczenskiBE~ and define
 \eqn\expandr{r(U)=r_{\rm st}(U)+\delta r(U)\,,}
where $r_{\rm st}$~ stands for the static configuration \lenden~.
Considering infinitesimal $\omega^2$,~ $\delta\rho$~  we can
expand \eomr~ and get a first order differential equation in
$\delta r'(U)$
\eqn\diffeqr{(U^3-U_0^3)(U^3-U_h^3)\delta r''+ {3 \delta r' \over
2U}(U^3(U_0^3-U_h^3)+2U^6-U_0^3 U_h^3)+ {\omega^2 r_{{\rm st}}(U)
\over 1- \omega^2 r_{{\rm st}}^2(U)}R^3 U^3 =0\,,}
which can be immediately integrated
 \eqn\deltarpri{ \delta r'(U)={U^3 \over
(U^3-U_0^3)^{3\over 2}(U^3-U_h^3)^{1\over 2}} \left( -R^3 \omega^2
\int_{U_{{\rm ref}}}^U {\sqrt{x^3-U_0^3} \over
\sqrt{x^3-U_h^3}}{r_{\rm st}(x) \over 1- \omega^2 r_{\rm
st}(x)^2}dx\right)\,.}
The integration constant has been encoded in $U_{{\rm ref}}$, that
we define as the value of $U$ at which $\delta r'(U)=0$. The
boundary conditions are that both ends of the string reach both
flavour branes orthogonally \bcs
\eqn\bound{r'(U_{D6}^{(i)})=r'_{\rm st}(U_{D6}^{(i)})+\delta
r'(U_{D6}^{(i)})=0\,.}
As we are in the long string regime, $U_0 \approx U_h$~and the
integral in \deltarpri~ is dominated by the vertical part of the
string. Moreover, unless ${M_q \over mq} \to \infty$, one can
safely $U_{{\rm ref}} \approx U_0$ as the point where the
derivative does not get modified must be in the horizontal region
of the string. In the degenerate case $m_q=M_q$, one has just
$U_{{\rm ref}} = U_0$ because of the symmetry. The relevant
integral is, then $\int_{U_0}^{U_{D6}} {r_{\rm st}(x) \over 1-
\omega^2 r_{\rm st}(x)^2}dx \approx {r \over 1-\omega^2
r^2}(U_{D6}-U_h)\,.${} From \bound, one gets two
conditions\footnote{$\dag$}{Notice, however, that the procedure to
obtain the constraints (3.18) differs from that in \KruczenskiME.
In \KruczenskiME, the string was cut into three pieces and (3.18)
is the condition for a proper gluing of the three pieces. Here,
the string was treated as a whole and (3.18) comes from the
boundary condition of the ends of the string hitting the branes
orthogonally.}
\eqn\omegar{1-\omega^2 r_{m_q}^2= T_s^{-1}\,\omega^2
r_{m_q}\,\,\tilde m_q\,,\quad  1-\omega^2 r_{M_q}^2=
T_s^{-1}\,\omega^2 r_{M_q}\,\,\tilde M_q\,,}
where $T_s$ is the tension of a string stretched at $U=U_h$
\eqn\tension{ T_s=\left({U_h \over R}\right)^{{3 \over 2}}}
 and
we have defined for each quark
\eqn\almostmass{ \tilde m ={(U_{D6}-U_h) U_{D6}^3 \over
U_{D6}^3-U_h^3}\,\,.}
Notice that for $U_{D6}\gg U_h$, one has $\tilde m \approx m$
\qmass, {\it i.e.} the mass of the quark, recovering the results
of \KruczenskiME, which are the holographic realization of the
hadron model of a relativistic string rotating in flat space
\refs{\ida,\ChodosGT}. Identifying $\tilde m, \tilde M$ with the
mass of the quarks, \omegar~ ensures the conservation of the
center of mass motion for a rotating string with massive endpoints
attached \ida
\eqn\cmass {{m_q r_{m_q} \omega \over \sqrt{1 - r_{m_q}^2
\omega^2}}- {T_s \over \omega}  \sqrt{1 - r_{m_q}^2 \omega^2} =
{M_q r_{M_q} \omega \over \sqrt{1 - r_{M_q}^2 \omega^2}}- {T_s
\over \omega} \sqrt{1 - r_{M_q}^2 \omega^2}\,.}

\medskip

This is certainly signaling that gluons inside the flux tube play
a {\sl dynamical} role and not only act as a mere geometric
constraint, contributing to the total energy of the system. If
not, momentum conservation would be just (consider $r_{m_q}^2
\omega^2\ll 1$)
 \eqn\msimple { m_q r_{m_q} = M_q r_{M_q} \,,}
as in the supersymmetric case. In this nonrelativistic limit, the
momentum conservation \cmass~ reads
\eqn\mlarge{ m_q r_{m_q} + {T_s\over 2} r_{m_q}^2 = M_q r_{M_q} +
{T_s\over 2} r_{M_q}^2\,.}
We expect a smooth transition between both behaviours,
\msimple,~and \mlarge.

\ifig\jjee{ Chew-Frautschi plot obtained numerically for $U_h=.2$,
$U_{D6^{(1)}}= 2\pi$ and $U_{D6^{(2)}}= 2\pi$ (squares),
$U_{D6^{(2)}}= 4\pi$ (triangles). The solid lines obtained using
\Eapprox, \Japprox, \omegar~ are the extrapolation of these curves
to the large-$J$ region. {}For comparison in the r.h.s. we have
plotted the experimental nucleon (full curve) \LyubimovKM~ and
Pomeron (dashed) \BrandtGI~ Regge trajectories.}{\epsfxsize
2.2in\epsfbox{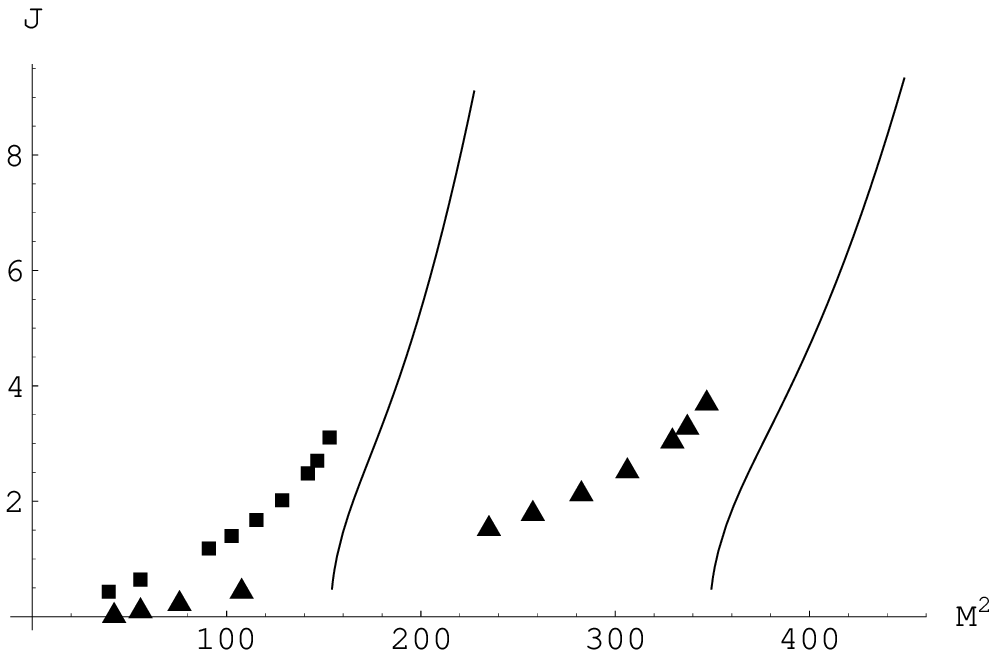} \epsfxsize 2.2in\epsfbox{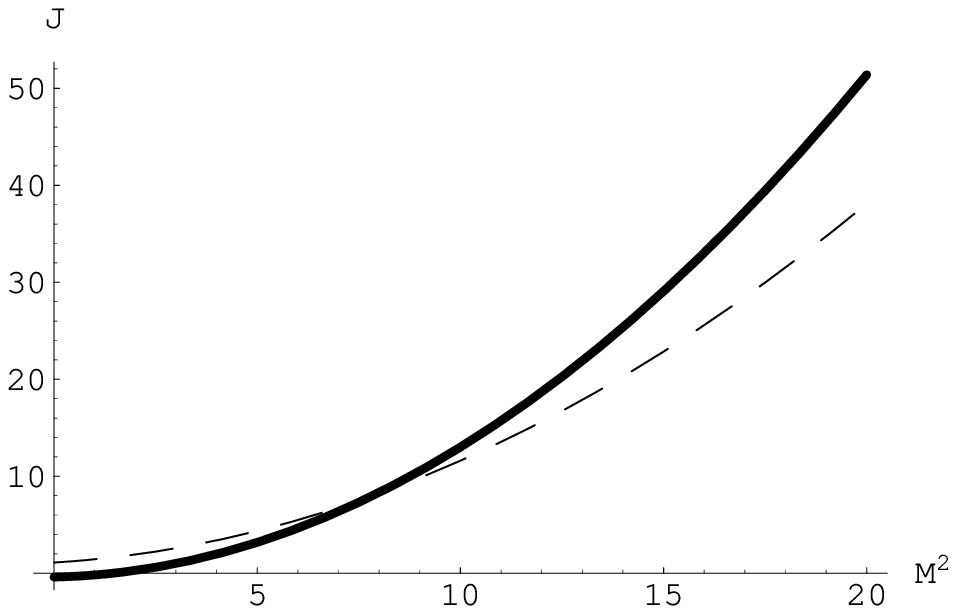}}

In \jjee , we have depicted the Chew-Frautschi corresponding to
this model. The curves giving the large-$J$ limit are represented
along with some points obtained numerically from \eomr, \eomrho,
\enqcd, \sqcd~ in the region of intermediate $J$. Unfortunately,
we have not been able to perform a trustable numerical analysis of
the very large-$J$ region, because when the string has a nearly
vertical-horizontal-vertical profile, numerical instabilities
appear. Despite the long extrapolation, the numerical results are
robust and trustable in the intermediate range of $J$ we
considered. {}For comparison in the r.h.s., we have presented some
experimental data. Notice that the qualitative similitude with the
l.h.s. in the full region is remarkable.

\bigskip

Given the above naive picture it is tempting to think in a very
special limit of QCD, heavy quark effective theory, consisting of
heavy hadrons composed by one heavy quark and one light one
\IsgurVQ. The heavy quark momenta is given by $P_Q^\mu=m_Q
v_Q^\mu+k^\mu\,,$~been $k^\mu$ a residual momenta of order
$\Lambda_{QCD}\,,$ while the hadron total momenta is $P_H^\mu =
m_H v^\mu$. In the heavy quark limit, the velocity superselection
rule leads that the residual momenta is negligible and all the
hadron momenta is carried by the heavy quark, $P_H\approx
P_Q$~\GeorgiUM. This setup is certainly susceptible to be obtained
by the above model using \mlarge\ via the identification
\eqn\heavy{ k^\mu \to m_q r_{m_q} + {T_s\over 2} r_{m_q}^2\approx
{\cal O}(\Lambda_{QCD})\,.}
Notice that \heavy~is suggesting that for a given mass the average
position of the light-quark with respect to the heavy one is not
arbitrary.

\bigskip
\noindent {\it 3.1.2. Intermediate $J$ strings}
\medskip

In the region of intermediate $J$, the holographic model differs
from the string rotating in flat space
\refs{\KruczenskiME,\ida,\ChodosGT}~ because the string profile
cannot be approximated any more by vertical and horizontal
regions. It may be regarded as a string of nonconstant tension
rotating in flat space. Similar models where considered in
\burakovsky~ in the search of more realistic models. Here, the
deviations from the constant tension are given uniquely by the
dynamics of the rotating string of the ten-dimensional theory. In
\jjee, the points obtained numerically show the deviation of the
formulae computed in the large angular momentum approximation
 when $J$ goes to intermediate values.

It is tempting to consider these deviations as the holographic
realization of the corresponding ones in QCD. However, one should
keep in mind that out of the large-$J$ limit, quantum corrections
to the semi-classical model must also be taken into account. These
kind of corrections where computed in \loopref~ for a static
configuration. Our setup is much more involved so it seems
hopeless to perform the analogous computation.

\bigskip
\noindent {\it 3.1.3. Small $J$ strings}
\medskip

For completeness, let us consider the $J \to 0$ limit, {\it i.e.}
rotating short strings. However, when the string is short, it
probes the KK direction, so one cannot think of this setup as the
holographic small-$J$ dual of QCD$_{3+1}$. By taking $r \to 0$ in
\eomr~ one finds
 \eqn\smallJ {r''- {3 \over 2 U} {U_h^3 -2 U^3
\over U_h^3 - U^3} r'-{R^3 \omega^2 \over U_h^3 - U^3} r=0\,.}
 We
have not been able to find an explicit solution for this equation.
However, a numerical computation shows that the qualitative
features are similar to those of section 2.1.1. for the
supersymmetric case (see  \ww~).

\bigskip

\subsec{ Quark-antiquark energy}

\ifig\EvsL{The quark-antiquark potential as a function of the
distance for  flavoured branes with, from bottom to top, $m_q=1$
and $M_q=1,2,3$. We also show the asymptotic limits (3.30), (3.32)
and (3.34). The r.h.s. shows for comparison the unquenched lattice
results \BaliVR. The squares denote the ground state static
potential while triangles stand for the first excited state.}
{\epsfxsize 2.2 in\epsfbox{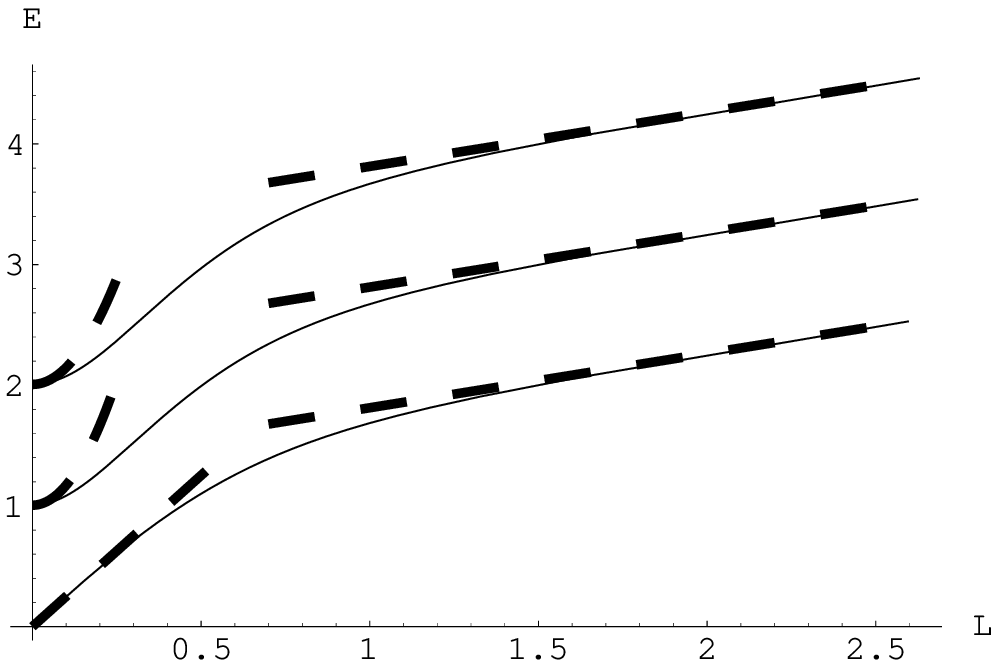} \epsfxsize 2.2
in\epsfbox{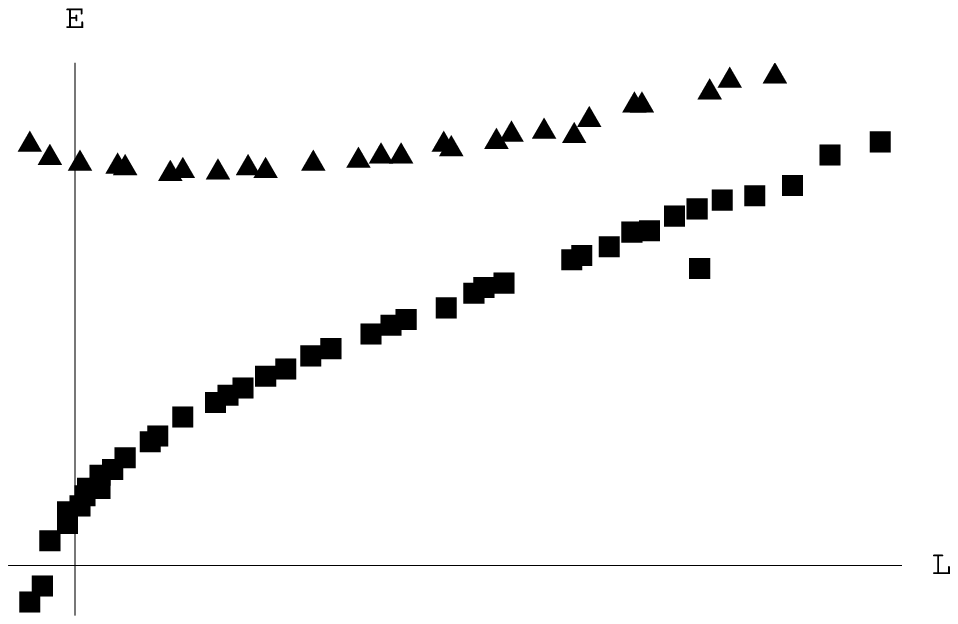}}

In this section we change tack, retreating from the semi-classical
limit of spinning strings to make qualitative statements about the
behavior of the static quark potential.

\smallskip

The D6 brane configuration has its world-volume along
$x_0,x_1,x_2,x_3$ and along two coordinates of the $S^4$. The
setup is realized by the string configuration
\eqn\wlsetup{ U=U(\sigma)\,,\quad X_0=\tau\,,\quad X_3=\sigma\,, }
where we have stuck to the static-gauge. The expressions for the
energy and the quark-anti quark separation are basically the same
as those obtained in \BrandhuberER\ but with an ultraviolet
cutoff, equal to the positions of the quarks on the D6-branes
 \eqn\logitud{ L= 2\, {R^{3/2}\over U_0^{1/2}}
\int_1^{U_{D6}^{(1)} \over U_0}\!dy\,{1\over
\sqrt{\left(y^3-1\right)\left(y^3-\eta ^3\right)}} +{R^{3/2}\over
U_0^{1/2}} \int_{U_{D6}^{(1)}\over U_0}^{U_{D6}^{(2)}\over
U_0}\!dy\,{1\over \sqrt{\left(y^3-1\right)\left(y^3-\eta
^3\right)}}\,, }
\eqn\energia{E={U_0\over \pi} \int_1^{U_{D6}^{(1)}\over
U_0}\!dy\,{y^3\over \sqrt{\left(y^3-1\right)\left(y^3-\eta
^3\right)}} + {U_0\over 2 \pi} \int_{U_{D6}^{(1)}\over
U_0}^{U_{D6}^{(2)}\over U_0}\!dy\,{y^3\over
\sqrt{\left(y^3-1\right)\left(y^3-\eta ^3\right)}}\,,\quad
\eta={U_h\over U_0}\,.}
In \EvsL~ we display the relation energy versus distance in the
$q\bar{q}$ system for a given pair
$\left(U_{D6}^{(1)}\,,U_{D6}^{(2)}\right)$. The gross features for
the ground state, convexity, match those of QCD \BachasXS\ but
with a relevant salient point: the existence of a linear behavior
at large distances (see discussion below). As in the D3-D7 case
the mass splitting only shifts upwards the curve $E$ vs. $L$ in
the large $L$ region without introducing any other change. At
high-energy the nondegenerate case shows once more a parabolic
behaviour. In fact the profile of the l.h.s. plot is identical to
the one in the supersymmetric case.

\bigskip
\noindent {\it 3.2.1. Mass Subtraction}
\medskip

As in the supersymmetric case the energy \energia\ is not a
renormalised expression once the integral upper cutoff go to the
boundary, and needs subtraction. With the same procedure as
previously \energia\ reads
\eqn\energiabis{E_{\rm ren} = {U_0\over \pi}
\int_1^{\infty}\!dy\,{y^3\over
\sqrt{\left(y^3-1\right)\left(y^3-\eta^3\right)}} - {U_0\over \pi}
\int_\eta^{\infty}\!dy\, \sqrt{y^3\over y^3-\eta^3}\,. }
Which, incidentally, does not match with the results of
\BrandhuberER.

\smallskip
As it is shown in \KruczenskiUQ\ , and unlike in the conformal
case, now the D6-brane is not located at a fixed point in the
transverse $U$ variable, but rather bends outwards in the region
where it is closest to the $D4$ stack of branes. We have checked
that the $q\bar{q}$ system minimizes its total energy when located
at the place in the D6-brane that minimizes its distance with
respect to the $D4$. This result coincides with the conclusions in
\KruczenskiUQ. In addition, as previously, in order to obtain a
light, finite $q\bar{q}$ system we have to make use of \energia\
instead of \energiabis.

\bigskip
\noindent {\it 3.2.2. Long and short strings}
\medskip

We turn to discuss the behaviour of long and short strings.
Notice, however, that although the relative distance between the
location of the $q\bar{q}$ pair on the D6-brane and the horizon
can be small, this by no means leads to the conclusion that the
distance between $q$ and $\bar{q}$ must be short as one can indeed
verify from \logitud\ by approaching the horizon at fixed
$U_{D6}^{(i)}$, {\sl i.e.} $y\rightarrow 1$. In fact for $\eta,y
\rightarrow 1$~ \logitud\ develops a single pole that can not be
cured, and hence the string approaches an infinite length.

\medskip
\noindent {\sl  Long strings}
\smallskip

A feature matching the expected QCD behavior is found at
low-energy. There the $q\bar{q}$ system has been hadronised and
from the macroscopic point of view we deal with hadrons. The slope
of these strings is constant. This is consistent with the  Regge
behaviour shown for rotating long strings in section 3.1.1. In
order to obtain long strings at a given finite distance for
$U_{D6}^{(i)}\gg \,(i=1,2)$ the string turning point should
approach the horizon. In that case $\eta = 1 + \epsilon$ and the
relation between  the energy and the length along the brane takes
the form
\eqn\firstlim{ {E} = {1\over 2\,\pi} \left( U_h\over
R\right)^{3/2} L + {1\over 2\,\pi}
\left(U_{D6}^{(1)}+U_{D6}^{(2)}-2 U_h
\right)\,. }
Notice that the inclusion of heavy masses shifts the intercept but
does not modify the slope, thus we still deal a Regge behaviour as
in the case of static quarks \BrandhuberER. The string tension is
the same as the one obtained in \BrandhuberER, and matches the
expression obtained from the energy formula for rotating strings
\PonsDK. The reason is that at large distances the main
contribution to the integrals \logitud , \energia\ comes from the
vicinity of the horizon, so the result is insensitive to the
presence of the cutoff at $U_{D6}^{(i)}$.

\medskip
\noindent {\sl Short strings}
\smallskip

As in the supersymmetric case, when the QCD string is very short,
its holographic dual is given by a string directly stretching
between both flavour branes, without bending (see \ddd). Then, the
expressions for the distance between the ends of the string and
its energy are just
 \eqn\shortnonsusy{ L= {R^{3/2}\over
U_0^{1/2}} \int_{U_{D6}^{(1)}\over U_0}^{U_{D6}^{(2)}\over
U_0}\!dy\,{1\over \sqrt{\left(y^3-1\right)\left(y^3-\eta
^3\right)}}\,,\ \ E= {U_0\over 2 \pi} \int_{U_{D6}^{(1)}\over
U_0}^{U_{D6}^{(2)}\over U_0}\!dy\,{y^3\over
\sqrt{\left(y^3-1\right)\left(y^3-\eta ^3\right)}}\,.}
Clearly, the $L\to 0$ corresponds to taking the integration limit
$U_0 \to 0$ (notice that this makes sense since in this case $U_0$
does not have the physical meaning of the string turning point).
Then, as in the supersymmetric case, one can check that, at
leading order, the energy has a parabolic dependence on the
distance
\eqn\nonsusypar{E=C_1 + C_2 L^2 + \cdots\,,}
where the constants are given by
\eqn\constants{C_1={1\over 2\pi}
\int_{U_{D6}^{(1)}}^{U_{D6}^{(2)}} dx{x^{3 \over 2} \over
\sqrt{x^3 - U_h^3}}=M_q-m_q\,,\,\,\,\,\,\,\,\,\, C_2={1\over 2\pi
R^3} \left(\int_{U_{D6}^{(1)}}^{U_{D6}^{(2)}} dx{1 \over x^{3
\over 2}\sqrt{x^3 - U_h^3}}\right)^{-1}\,.}
{}For the sake of completeness, let us mention that, in the
degenerate case, this reasoning is clearly not valid and the
string must stretch with bending. Then, the leading order
expansion leads to a constant slope behaviour for the energy
\eqn\seconlim{E =  {1\over 2\,\pi} \left({U_h\over
R}\right)^{3/2}L + \cdots \,.}
This shape has been quite long discussed in QCD, see for instance
\ChernodubBK. With the range of energy explored in the actual
lattice data this effect can be interpreted as a renormalon
artefact and has no further physical consequences. In the present
case we can understand it in an energy basis: in the coordinates
\metricM\ $E \sim U$. As we increase the $U$ coordinate also the
energy is increased. At certain energy, $E\sim M_{KK}\sim {2\over
3} {R^{3/2}\over U_0^{1/ 2}}$ there is an opening of the dimension
on the $S^1$, thus we deal with QCD$_{4+1}$ instead of
QCD$_{3+1}$.

To substantiate this claim in an analytic way, we have performed
the limit $U_{D6}^{(i)} \rightarrow \infty$, and renormalised the
energy with the subtraction \energiabis. We can observe that for
small distances between the quarks, the energy behaves as
$$
E \approx - \eta\, U_0^{1/2} {1\over L^2} + {\rm ct}. \,,
$$
thus showing a Coulomb phase for $QCD_{4+1}$.

\subsec{Quark-antiquark potential screening}

\lref\BornIV{ K.~D.~Born, E.~Laermann, N.~Pirch, T.~F.~Walsh and
P.~M.~Zerwas, ``Hadron Properties In Lattice QCD With Dynamical
Fermions,'' Phys.\ Rev.\ D {\bf 40}, 1653 (1989).
}


\lref\BrodskyNB{ S.~J.~Brodsky, S.~Menke, C.~Merino and
J.~Rathsman, ``On the behavior of the effective QCD coupling
alpha(tau)(s) at low scales,'' Phys.\ Rev.\ D {\bf 67}, 055008
(2003) [arXiv:hep-ph/0212078].
}

\lref\ZwanzigerCF{ D.~Zwanziger, ``Non-perturbative Faddeev-Popov
formula and infrared limit of QCD,'' Phys.\ Rev.\ D {\bf 69},
016002 (2004) [arXiv:hep-ph/0303028].
}

\lref\MattinglyEJ{ A.~C.~Mattingly and P.~M.~Stevenson,
``Optimization of R(e+ e-) and 'freezing' of the QCD couplant at
low-energies,'' Phys.\ Rev.\ D {\bf 49}, 437 (1994)
[arXiv:hep-ph/9307266].
}

\lref\BornCQ{ K.~D.~Born, E.~Laermann, R.~Sommer, P.~M.~Zerwas and
T.~F.~Walsh, ``The Interquark potential: A QCD lattice analysis,''
Phys.\ Lett.\ B {\bf 329}, 325 (1994).
}

\lref\HellerRZ{ U.~M.~Heller, K.~M.~Bitar, R.~G.~Edwards and
A.~D.~Kennedy, ``The Heavy quark potential in QCD with two flavors
of dynamical quarks,'' Phys.\ Lett.\ B {\bf 335}, 71 (1994)
[arXiv:hep-lat/9401025].
}

It is instructive to gain some insight on the effects of including
the dynamical quarks, mimicked for the D6-brane. In \ELplot ,
r.h.s., we have plotted (thick curve) the quark-antiquark energy
obtained by taking $U_{D6}\to\infty$ and subtracting the bare mass
\energiabis. As one can see, already keeping the effective field
theory in (3+1)d, the effect of introducing dynamical quarks is
quite noticeable and with similar pattern to the supersymmetric
case. This change of behaviour can be understood as a screening
effect: while in a pure glue dynamic system composed of two
infinite external heavy-quarks the interaction can be only
mediated via gluons, in a system with fundamental matter the gluon
propagator itself is {\sl dressed} with quarks. This fact weakness
the interaction between the valence-quarks and the potential wall
gets broadest. To manifest this effect we have fitted to \EvsL ,
inside the (3+1)d region, two phenomenological models for the case
$m_q = M_q= 1$. The first model is given by the screened potential
\BornIV\
\eqn\lattscr{ V(r) = \left( -{\alpha \over r} + \sigma r\right)
{1-e^{-\mu r} \over \mu\, r}\,, }
and the second by the well-known Cornell potential
\eqn\cornell{ V(r) = V_0 + b r - {a\over r}\,. }
In this case the model realised with D6-branes behaves most as a
screened system at short distances. Although the relative
different between potentials is so small that no conclusive
statement on the behaviour of the system  can be obtained. Similar
conclusions are achieved by lattice studies
\refs{\HellerRZ,\BornCQ}.

As a concluding remark on this section we mention that for both,
the singlet and the first excited state, see the UV region in
\EvsL , the beta function for the D4-D6 system has an UV fixed
point. This is in the singlet case is in clear contrast with our
actual knowledge on perturbative QCD.

\subsec{'t Hooft line: monopole-monopole interaction}

\lref\MantonER{ N.~S.~Manton, ``The Force Between 'T
Hooft-Polyakov Monopoles,'' Nucl.\ Phys.\ B {\bf 126}, 525 (1977).
}
\lref\HananyIE{ A.~Hanany and E.~Witten, ``Type IIB superstrings,
BPS monopoles, and three-dimensional gauge dynamics,'' Nucl.\
Phys.\ B {\bf 492}, 152 (1997) [arXiv:hep-th/9611230].
}
\lref\DiaconescuRK{ D.~E.~Diaconescu, ``D-branes, monopoles and
Nahm equations,'' Nucl.\ Phys.\ B {\bf 503}, 220 (1997)
[arXiv:hep-th/9608163].
}
\lref\GrossGK{ D.~J.~Gross and H.~Ooguri, ``Aspects of large N
gauge theory dynamics as seen by string theory,'' Phys.\ Rev.\ D
{\bf 58}, 106002 (1998) [arXiv:hep-th/9805129].
}

A complementary way of looking at confinement is via the magnetic
dual of the quark, the monopole. The configuration realizing a
magnetic monopole from \metricM\ is that of a D2-brane ending on
the initial D4-brane. The monopole is obtained by wrapping the
D2-brane on the compactified $S^1$ \refs{\HananyIE,\DiaconescuRK}.
With our conventions, the mass of a free monopole is
\eqn\monopolem{m_m = K\, (U_{(D6)}-U_h)\,,} where we have defined
the constant $K\equiv 4 \pi T_{D2} R^{3/2}  U_h^{-1/2}g_s^{-1}$.
The potential between the monopoles is obtained by placing two
 monopoles  at distance $L$ in
the $x_1$-axis. Consider a D2-brane  parametrised in term of
$(t,U(x_1) , \theta_2)$. The induced metric on the brane is then
\GrossGK\
 $$g_{tt}= -\left({U\over R}\right)^{3/2}
\,,\quad
g_{UU}=
\left({R\over U}\right)^{3/2}f(U)^{-1}(1+\left({U\over R}\right)^3
f(U) x_1'(U)^2)\,,$$
$$
\quad g_{\theta_2 \theta_2}= \left({U\over
R}\right)^{3/2}{4 \over 9}{R^3 \over U_h} f(U)\,.
$$
Inserting it in the D2 action \eqn\ddosaction{E_{\rm m-m}=
T_{D2}\int dx_1 d\theta_2\, e^{-\phi}
\sqrt{g_{tt}\,g_{UU}\,g_{\theta_2 \theta_2}}\,, } we can extract
the distance between the ends of the flux-tube and the energy
stored in it (for simplicity, we consider monopoles with
degenerate mass)
\eqn\mlenght{L= 2 {R^{3/2} \over U_h^{1/2}} \int_1^{{U_{D6}\over
U_0}} {\sqrt{\eta\left(1-\eta^3\right)}\over
\sqrt{\left(y^3-1\right) \left(y^3-\eta^3\right)}}\,, }
\eqn\hamiltonian{E = K\,
U_0 \int_1^{{U_{D6}\over U_0}} dy\, \sqrt{{y^3-\eta^3\over
y^3-1}}\,, }
where $\eta$ is defined as in \energia.

{F}or small distance there is a configuration of the flux tube
between the monopoles for which the energy \hamiltonian~ is less
than that of two free monopoles \monopolem. As we increase the
distance between the monopoles the energy stored in the tube grows
up to certain $L_{\rm crit}$ where it reaches $2 m_m$ signaling
the breaking of the flux tube. {F}rom these the most stable
configuration corresponds to two flux tubes stretching from each
of the monopoles in the D6-brane up to the horizon. This leads to
a constant energy density in accordance with field theory
expectations \MantonER.

Notice that for probe branes near the horizon, light-quark masses,
the breaking takes place at lower values of $L_{\rm crit}$, while
for probe branes supporting heavy quarks the breaking occurs at
higher ones.

\newsec{Summary}

By considering open rotating strings attached to probe branes we
have evaluated semi-classically some properties of a system
mimicking excited mesons formed by different constituents masses.
We focused on two basic aspects: {\sl i)} the Chew-Frautschi plot
and {\sl ii)} the $q\bar{q}$ system energy. The first quantity is
obtained by introducing open rotating strings with ends attached
to an embedded probe brane an extending into a transverse
coordinate. While the second is a static quantity. In the actual
range of energies the QCD static potential has the form
$$
V_{q\bar{q}}(r)=  -{e\over r}+ c r\,.
$$
Adding light flavours we move away of this limit and new effects,
as the spin, play an essential role. These potentials are reliable
to study heavy quark interaction, but when light quarks are
involved, relativistic effects are important and even the concept
of inter-quark {\sl potential} remains questioned.

Bearing this lack in mind we construct a system composed by two
different, tuneable masses by adding two probe branes and
separating them in a plane orthogonal to the initial brane stack
world-volume.

The first case of backgrounds we treated is a supersymmetric
setup, the D3-D7 brane system, generalising the extremal case
\KruczenskiBE. At low-energy neither the $J-E$ relation nor the
$E-L$ show a qualitative difference with respect to the degenerate
case but at moderate-high energy differences are substantial.
Notice that at qualitative level, and without any apparent reason,
the $J-E$ and $E-L$ plots show the same shape, see \ejdos,
\ELplot.

The second studied background is non-supersymmetric and is
obtained considering a thermal cycle in 10d. The gross features of
the Chew-Frautschi plot for hadronised systems resembles that of
heavy-quarkonia in QCD. The $E-L$ relation presents a Regge type
behaviour for large separations, matching the field theory
expectations. At high-energy there is however also a Regge
behaviour in the case of degenerate masses. This result can be
interpreted, {\sl granus salis}, as the possible presence of a
screening effect as one can see by fitting the $q\bar{q}$ energy
to two different potentials. There is however a rather surprising
quantitative agreement with lattice findings if the singlet and
1st excited states are considered. The basic tenets of the model
can be tested by lattice Monte-Carlo simulations of higher excited
states. The last remark we have made on confinement concerns the
picture of a monopole-monopole condensation, which fulfills the
field theory expectations.

\lref\SakaiCN{ T.~Sakai and S.~Sugimoto, ``Low energy hadron
physics in holographic QCD,'' arXiv:hep-th/0412141.
}

\medskip
We stress that the all the mentioned features are independent of
having implemented supersymmetry in the theory. In this respect we
believe that the aforementioned characteristics of the
Chew-Frautschi plot or in the quark-antiquark potential in the
D4-D8 \SakaiCN\ model must led to the same qualitative results as
those in the D4-D6. Nevertheless it would be interesting to see if
the existence, at large-energy, of a Regge behaviour persists in
this model.

\bigskip

{\bf Acknowledgements}

\smallskip
We would like to thank Ll.~Ametller, A.~Cotrone, C.~N\'u\~nez,
J.~M.~Pons and A.~Ramallo for comments. The work of A.~P. was
partially supported by INTAS grant, 03-51-6346, CNRS PICS $\#$
2530, RTN contracts MRTN-CT-2004-005104 and MRTN-CT-2004-503369
and by a European Union Excellence Grant, MEXT-CT-2003-509661. The
work of P.~T. is supported in part by the European Community's
Human Potential Programme under contract MRTN-CT-2004-005104
`Constituents, fundamental forces and symmetries of the universe'.

\listrefs
\bye